\documentclass[12pt]{article}

\usepackage[margin=1in]{geometry}

\usepackage{comment}

\usepackage{enumitem} 

\usepackage{centernot}

\usepackage{array} 
\usepackage{arydshln}

\usepackage{multirow}

\usepackage{setspace} 
\usepackage{caption}
\renewcommand{\arraystretch}{0.6} 

\usepackage{amsmath}
\usepackage{amssymb}
\usepackage{amsthm} 
\usepackage{mathtools} 

\usepackage[utf8]{inputenc} 
\usepackage[english]{babel}

\usepackage{graphicx} 

\usepackage{float}
\usepackage{placeins}

\newtheorem{assumption}{Assumption}

\newtheorem{condition}{Condition}

\usepackage[title]{appendix}

\numberwithin{equation}{section}

\newcommand\blfootnote[1]{
  \begingroup
  \renewcommand\thefootnote{}\footnote{#1}
  \addtocounter{footnote}{-1}
  \endgroup
}

\usepackage[hang]{footmisc}
\usepackage{hyperref}
\hypersetup{colorlinks=true, allcolors=black}

\usepackage[authoryear,round,sort,comma,longnamesfirst]{natbib} 
\usepackage{myapalike}
\begin{document}

\begin{titlepage}
\begin{center}
\linespread{1.2}
\Large{\textbf{A Gaussian smooth transition vector autoregressive model: An application to the macroeconomic effects of severe weather shocks}}\\

\vspace{0.5cm} 
\Large{Markku Lanne \quad Savi Virolainen}\\
\large{University of Helsinki}\\
\vspace{1.0cm}

\begin{abstract}
\noindent We introduce a new smooth transition vector autoregressive model with a Gaussian conditional distribution and transition weights that, for a $p$th order model, depend on the full distribution of the preceding $p$ observations. 
Specifically, the transition weight of each regime increases in its relative weighted likelihood. This data-driven approach facilitates capturing complex switching dynamics, enhancing the identification of gradual regime shifts. In an empirical application to the macroeconomic effects of a severe weather shock, we find that in monthly U.S. data from 1961:1 to 2022:3, the shock has stronger impact in the regime prevailing in the early part of the sample and in certain crisis periods than in the regime dominating the latter part of the sample. This suggests overall adaptation of the U.S. economy to severe weather over time.\\[0.5cm] 


\end{abstract}

\vfill

\blfootnote{The authors thank Pentti Saikkonen for the useful discussions and the Research Council of Finland for financial support (Grant 347986).}
\blfootnote{Contact address: Savi Virolainen, Faculty of Social Sciences, University of Helsinki, P. O. Box 17, FI–00014 University of Helsinki, Finland; e-mail: savi.virolainen@helsinki.fi. ORCiD ID: 0000-0002-5075-6821.}
\blfootnote{The authors have no conflict of interest to declare.}

\end{center}
\end{titlepage}

\section{Introduction}
Smooth transition vector autoregressive (STVAR) models that allow for gradual shifts in parameter values have become popular due to their ability to capture nonlinear dynamics in time series data \citep[for a survey of the literature, see, e.g.,][and the references therein]{Hubrich+Terasvirta:2013}. They extend the conventional linear vector autoregressive (VAR) model by allowing for smooth transitions between regimes, each of which is characterized by a linear vector autoregression with different coefficients or error covariance matrices. Each observation is the sum of a weighted average of the conditional means of the regimes and a random error, whose covariance matrix is a weighted average of the covariance matrices of the regimes. 
Like other nonlinear structural VAR models, the structural counterparts of STVAR models enable tracing out the causal effects of economic shocks, which may depend on the initial state of the economy or on the sign or size of the shock.


Different STVAR models are obtained by specifying the transition weights or the error distribution in various ways. In this paper, we introduce the Gaussian smooth transition vector autoregressive (GSTVAR) model, where the transition weights are similar to the mixing weights of the Gaussian mixture vector autoregressive (GMVAR) model of \cite{Kalliovirta+Meitz+Saikkonen:2016}. For a $p$th order model, the transition weights in both models depend on the full distribution of the preceding $p$ observations, which enables capturing  complicated switching dynamics. Specifically, the greater the relative weighted likelihood of a regime is, the greater its transition weight is, which facilitates associating statistical characteristics and economic interpretations to the regimes. However, in contrast to the GMVAR model, which incorporates unobserved discrete regime switches, our GSTVAR model has the advantage that it allows for capturing gradual shifts in the dynamics of the data. Moreover, the identified shocks of the structural GSTVAR model can  be recovered from the fitted model and used in impulse response analysis to further reflect the properties of the data, which is not the case with the structural GMVAR model.

Probably the most popular among the STVAR models put forward in the previous literature is  the logistic STVAR (LSTVAR) model \citep{Anderson+Vahid:1998} characterized by logistic transition weights. In the LSTVAR model, the interpretation of the regimes is clear, and it can accommodate exogenous as well as endogenous switching variables. These properties are shared by the often employed threshold VAR (TVAR) model \citep{Tsay:1998}, albeit the latter features discrete regimes instead of allowing for smooth transitions between them. Hence, these models are well-suited, when the transition weights can be expected to depend on the level of some specific switching variables. In contrast, our GSTVAR model facilitates capturing more complicated switching dynamics and results in more clearly data-driven regimes, but it may not always be straightforward to interpret the regimes economically.

Besides STVAR and TVAR models, there are a number of alternative approaches to modeling nonlinearities in multivariate time series, including so-called Markov-switching VAR (MS-VAR) models \citep[][see also \citealp{Hamilton:1989, Hamilton:1990}]{Krolzig:1997} and time-varying parameter VAR (TVP-VAR) models (\citealp{Cogley+Sargent:2001}; \citealp{Cogley+Sargent:2005}; \citealp{Primiceri:2005}; \citealp{Koop+Leon-Gonzalez+Strachan:2009}; among others). The former incorporate discrete regime switches that are random and unobserved, and the switching probabilities depend on only the preceding regime. While they can flexibly capture nonlinearities in the data, unlike our GSTVAR model, they do not accommodate gradual shifts in the dynamics nor complicated switching dynamics. As to TVP-VAR models, compared to our GSTVAR model, they have the major limitation that the stochastic processes governing the changes in parameter values are exogenous to the variables included, which may lead to overlooking important endogenous dynamics.


We apply the structural GSTVAR model to find out about the macroeconomic effects of severe weather. This issue was recently studied by \cite{Kim+Matthes+Phan:2022} by a STVAR model with transition weights defined by a linear time trend, which only allows a single switch from one regime to the other. Following them, we include an indicator of severe weather in addition to a number of macroeconomic variables in the model and estimate it on monthly U.S. data from 1961:1 to 2022:3. In line with \cite{Kim+Matthes+Phan:2022}, we find evidence in favor of nonlinearity, and a GSTVAR model with two regimes is deemed sufficient. However, our regimes are characterized by features quite different from theirs. One regime dominates the earlier part of the sample period, particularly the volatile periods of the 1970s and 1980s, and also prevails later during the Financial crisis and the COVID-19 crisis, while the latter part of the sample period is predominantly characterized by the other regime. 

In both regimes, a recursively identified positive severe weather shock decreases GDP, consumer prices, and the interest rate, but the effects are clearly stronger in the regime prevailing in the earlier part of the sample period and in certain crisis periods. 
In contrast to \cite{Kim+Matthes+Phan:2022}, who found the effects of severe weather stronger in the latter part of the sample period, we thus conclude that the U.S. economy has adapted to the changing distribution of weather related shocks over time. However, since strong effects are found also in the crisis periods, it seems that the apparent adaptation does not provide sufficient resilience when the economy is in a vulnerable state. 

The rest of this paper is organized as follows. Section~\ref{sec:stvar} first presents our framework for STVAR models and then discusses their stationarity and ergodicity. We also introduce the structural STVAR model and some tools useful in empirical analysis based on it. Finally, the new Gaussian STVAR (GSTVAR) model is introduced by specifying the transition weights and the error distribution. In Section~\ref{sec:estimation}, model selection and estimation of the parameters by the method of maximum likelihood are discussed. Section~\ref{sec:empapp} contains the empirical application to the macroeconomic effects of severe weather shocks. Section \ref{sec:conclusion} concludes the paper. Further details can be found the in the appendices, and the introduced methods are implemented to the accompanying R package sstvars \citep{sstvars}, available in the CRAN repository.

\section{Smooth transition vector autoregressive models}\label{sec:stvar}
\subsection{General framework for STVAR models}\label{sec:genstvar}

The STVAR model with $M$ regimes and autoregressive order $p$ considered in this paper can be written as
\begin{align}
y_t &=\sum_{m=1}^M \alpha_{m,t}\mu_{m,t} + u_t, \quad u_{t} \sim (0, \Omega_{y,t}),\label{eq:stvar1} \\
\mu_{m,t} &= \phi_{m,0} + \sum_{i=1}^{p}A_{m,i}y_{t-i}, \quad m=1,...,M,\label{eq:stvar2}\\
\Omega_{y,t} &= \sum_{m=1}^M \alpha_{m,t}\Omega_m, \label{eq:stvar3}
\end{align}
where $\phi_{1,0},...,\phi_{M,0}\in\mathbb{R}^{d}$, $m=1,...,M$, are the intercept parameters, $A_{1,i},...,A_{M,i}\in\mathbb{R}^{d\times d}$, $i=1,...,p$, are the autoregressive matrices, and $\Omega_1,...,\Omega_M$ are the positive definite $(d\times d)$ covariance matrices of the regimes. The serially uncorrelated reduced form innovations $u_t$ follow some distribution with zero mean and conditional covariance matrix $\Omega_{y,t}$.

The transition weights $\alpha_{m,t}$ are assumed to be $\mathcal{F}_{t-1}$-measurable functions of $\lbrace y_{t-j}, j=1,...,p \rbrace$ and to satisfy $\sum_{m=1}^{M}\alpha_{m,t}=1$ at all $t$. These weights convey the relative proportions of the regimes at each point in time and determine how the process shifts between them. Specifically, when the process is completely in one of the regimes, the transition weight $\alpha_{m,t}$ of that regime equals unity, while the weights of the rest of the regimes are equal to zero. As the process begins a shift towards another regime, the transition weight of the emerging regime increases due to the dynamics captured in the transition weight function through the preceding observations. Concurrently, the weight of the previously dominant regime decreases. 

It is easy to see that, conditional on $\mathcal{F}_{t-1}$, the conditional mean of the above-described process is $\mu_{y,t} \equiv E[y_t|\mathcal{F}_{t-1}] = \sum_{m=1}^M \alpha_{m,t}\mu_{m,t}$, and its conditional covariance matrix is $\Omega_{y,t} = \text{Cov}(y_t|\mathcal{F}_{t-1}) = \sum_{m=1}^M \alpha_{m,t}\Omega_m$. That is, the conditional mean is a weighted sum the regime-specific means $\mu_{m,t}$ with the weights given by the transition weights $\alpha_{m,t}$, whereas the conditional covariance matrix is a weighted sum of the regime-specific conditional covariance matrices $\Omega_m$.   

For standard asymptotic inference, the STVAR process must be stationary and ergodic. To verify that this is indeed the case, we rely on the sufficient condition for stationarity and ergodicity derived by \cite{Kheifets+Saikkonen:2020} \citep[based on the more general results of][]{Saikkonen:2008}. However, their parametrization is not quite the same as ours, so their result must be slightly modified for our purposes (see Appendix~\ref{sec:statcond} for details). In terms of the companion form AR matrices of the regimes defined as
\begin{equation}\label{eq:boldA}
\boldsymbol{A}_m = 
\underset{(dp\times dp)}{\begin{bmatrix}
A_{m,1} & A_{m,2} & \cdots & A_{m,p-1} & A_{m,p} \\
I_d  & 0     & \cdots & 0            & 0 \\
0     & I_d  &             & 0            & 0 \\
\vdots &   & \ddots & \vdots    & \vdots \\
0     & 0     & \hdots & I_d         & 0
\end{bmatrix}}, \
m=1,...,M,
\end{equation}
\citet[][Theorem 1]{Kheifets+Saikkonen:2020} show 
that if the following condition holds, the STVAR process is ergodic stationary (both strictly and second-order). 
\begin{condition}\label{cond:sufficient}
$\rho(\lbrace \boldsymbol{A}_1,...,\boldsymbol{A}_M \rbrace) < 1$.
\end{condition}
Here 
\begin{equation}
\rho(\mathcal{A}) = \underset{j\rightarrow \infty}{\limsup}\left(\underset{A\in \mathcal{A}^j}{\sup}\rho(A) \right)^{1/j}
\end{equation}
denotes the joint spectral radius (JSR) of a finite set of square matrices $\mathcal{A}$ with $\mathcal{A}^j=\lbrace A_1A_2...A_j:A_i\in\mathcal{A}\rbrace$ and $\rho(A)$ is the spectral radius of the square matrix $A$. As \cite{Kheifets+Saikkonen:2020} note, Condition~\ref{cond:sufficient} is not necessary for ergodic stationarity of the process, meaning that if it does not hold, we just cannot use the result to state whether the process is ergodic stationary or not.

It is worth mentioning that a necessary condition for Condition~\ref{cond:sufficient} is that the usual stability condition is satisfied for each of the regimes, as $\max(\rho(\boldsymbol{A}_1),...,\rho(\boldsymbol{A}_M))\leq \rho(\lbrace \boldsymbol{A}_1,...,\boldsymbol{A}_M \rbrace)$ \citep[see][and the references therein]{Kheifets+Saikkonen:2020}. Therefore, the following condition, which is analogous to Corollary~1 of \cite{Kheifets+Saikkonen:2020}, is necessary for Condition~\ref{cond:sufficient}.
\begin{condition}\label{cond:necessary}
$\max\lbrace \rho(\boldsymbol{A}_1),...,\rho(\boldsymbol{A}_M)\rbrace<1$.
\end{condition}

Since the sufficient Condition~\ref{cond:sufficient} is computationally costly to verify with reasonable accuracy \citep[see, e.g.,][]{Chang+Blondel:2013}, it is useful to employ the more easily verified necessary Condition~\ref{cond:necessary} in numerical estimation (see  Section~\ref{sec:estimation}). As our estimation procedure produces a set of alternative local solutions, Condition~\ref{cond:sufficient} can be checked for each of them after the estimation, and the best local solution is selected among those for which stationarity can be verified. 

A number of methods for bounding the JSR have been proposed in the literature, many of which are discussed by \cite{Chang+Blondel:2013}. The accompanying R package sstvars \citep{sstvars} implements the branch-and-bound method of \cite{Gripenberg:1996}. The JSR toolbox in MATLAB \citep{Jungers:2023}, in turn, automatically combines various methods in the estimation of the JSR to enhance computational efficiency.

\subsection{Structural STVAR model} \label{sec:SSTVAR}

To conduct structural analysis, we need to find the structural counterpart of the STVAR model that involves orthogonal, serially uncorrelated structural errors, or shocks, $e_t$. This amounts to finding a non-singular ($d\times d)$ impact matrix $B_t$ such that the conditional covariance matrix of $e_t = B_t^{-1}u_t$, conditional on $\mathcal{F}_{t-1}$, is a diagonal matrix  (typically normalized to the identity matrix). In other words, $B_t$ must be such that $B_t^{-1}\Omega_{y,t}B_t'^{-1}=I_d$. However, without additional identifying restrictions $B_t$ satisfying this equation is not unique, and as discussed by, e.g., \citeauthor{Kilian+Lutkepohl:2017} (\citeyear{Kilian+Lutkepohl:2017}, Chapter 18), imposing such restrictions in nonlinear structural VAR models may not be straightforward. In our empirical application in Section \ref{sec:empapp}, we consider a recursive model, where $B_t$ is obtained by a lower-triangular Cholesky decomposition of the conditional covariance matrix $\Omega_{y,t}$.


Conventional impulse responses can be computed separately for each regime in the structural STVAR model, as has been done in much of the nonlinear structural VAR literature. However, by thus precluding future regime switches, this approach changes the structure of the model and makes the impulse response analysis subject to Lucas critique. Moreover, the expected effects of the structural shocks generally depend on the initial values of the variables as well as on the sign and size of the shock, which is not accounted for by the conventional impulse responses. An appropriate alternative is the generalized impulse response function (GIRF) of \cite{Koop+Pesaran+Potter:1996}, defined as
\begin{equation}\label{eq:girf}
\text{GIRF}(h,\delta_j,\mathcal{F}_{t-1}) = \text{E}[y_{t+h}|\delta_j,\mathcal{F}_{t-1}] - \text{E}[y_{t+h}|\mathcal{F}_{t-1}],
\end{equation}
where $h$ is the horizon and $\mathcal{F}_{t-1}=\sigma\lbrace y_{t-j},j>0\rbrace$ as before. The first term on the right side of (\ref{eq:girf}) is the expected realization of the process at time $t+h$ conditional on a structural shock of sign and size $\delta_j \in\mathbb{R}$ in the $j$th element of $e_t$ at time $t$ and the previous observations. The latter term on the right side is the expected realization of the process conditional on the previous observations only. The GIRF thus expresses the expected difference in the future outcomes when the structural shock of sign and size $\delta_j$ in the $j$th element hits the system at time $t$ as opposed to all shocks being random. An interesting feature of the structural STVAR model is that besides the generalized impulse response functions of the observable variables, it produces the GIRFs of the transition weights $\alpha_{m,t}$, $m=1,...,M$, which yield information on the dynamic effects of the shocks on regime switches. They are obtained by replacing $y_{t+h}$ with $\alpha_{m,t}$ on the right side of Equation~(\ref{eq:girf}). 

Typically STVAR models have a $p$-step Markov property, and this is also the case with our GSTVAR model to be introduced in Section \ref{sec:gstvar}. Thus conditioning on (the $\sigma$-algebra generated by) the $p$ previous observations $\boldsymbol{y}_{t-1}=(y_{t-1},...,y_{t-p})$ is effectively the same as conditioning on $\mathcal{F}_{t-1}$ at time $t$ and later. To estimate the GIRFs conditional on the economy being in a specific regime, say $\tilde{m}$, when the shock arrives, we select the length $p$ histories $\boldsymbol{y}_{t-1}$ from the data that indicate the dominance of this regime in the time period $t$. That is, we select the histories $\boldsymbol{y}_{t-1}$ for which the transition weight $\alpha_{\tilde{m},t}$ is large, say, larger than $0.75$, indicating that the process is largely in Regime~$\tilde{m}$. For each of these histories $\boldsymbol{y}_{t-1}$, we compute the GIRFs to the structural shock $e_{it}$ recovered from the data (the $i$th element of the estimate of $B_t^{-1}(y_t - \mu_{y,t})$), so they reflect the properties of the data. 
To then make GIRFs to shocks with different signs and sizes comparable, they are all scaled to correspond to some fixed (say, unit) instantaneous increase in one of the variables. Finally, the distribution of GIRFs obtained for all histories $\boldsymbol{y}_{t-1}$ can be superimposed in a so-called ``shotgun plot" \citep[cf.][]{Inoue+Kilian:2016}. When a low level of opacity is used, the darkness of each region in the figure indicates the concentration of GIRFs in it (for an example, see Figure \ref{fig:girfplot}). A detailed description of our Monte Carlo algorithm for estimating the GIRF is presented in Appendix~\ref{sec:montecarlo_girf}.\footnote{Another way to estimate GIRFs conditional on the economy being in a specific regime when the shock arrives is to generate the initial values $\boldsymbol{y}_{t-1}=(y_{t-1},...,y_{t-p})$ from the stationary distribution of that regime. Comparison of regime-dependent GIRFs for different signs and sizes of the shock then reveals asymmetries with respect to these features within the regime.}

Similarly to the conventional impulse response functions being unsuitable for impulse response analysis in the structural STVAR model (due to their inability to capture asymmetries in the effects of the shocks or to take future switches in the regime into account), the conventional forecast error variance decomposition is unsuitable for tracking the relative contribution of each shock to the variance of the forecast errors. Instead, the generalized forecast error variance decomposition (GFEVD) of \cite{Lanne+Nyberg:2016} can be used. It is defined for variable $i$, shock $j$, and horizon $h$ as
\begin{equation}
\text{GFEVD}(h,y_{it}, \delta_j,\mathcal{F}_{t-1}) = \frac{\sum_{l=0}^h\text{GIRF}(l,\delta_j,\mathcal{F}_{t-1})_i^2}{\sum_{k=1}^d\sum_{l=0}^h\text{GIRF}(l,\delta_k,\mathcal{F}_{t-1})_i^2},
\end{equation}
where $h$ is the chosen horizon and $\text{GIRF}(l,\delta_j,\mathcal{F}_{t-1})_i$ is the $i$th element of the related GIRF. The GFEVD is otherwise similar to the conventional forecast error variance decomposition but with GIRFs in the place of conventional impulse response functions, and it can be interpreted in a similar manner to the conventional forecast error variance decomposition.

\subsection{Gaussian STVAR model}\label{sec:gstvar}

Specifying a particular STVAR (and the related structural STVAR) model from the definition in Section~\ref{sec:genstvar} amounts to specifying the distribution of the reduced form innovations $u_t$ (or the structural shocks $e_t$) and the transition weights $\alpha_{m,t}$. In this paper, we propose a Gaussian STVAR model with standard normal distributions for the structural errors $e_t$. Hence, the conditional distribution of $y_t$, conditional on $\mathcal{F}_{t-1}$, is Gaussian and characterized by the density function
\begin{equation}\label{eq:gausconddist}
f(y_t|\mathcal{F}_{t-1}) = n_d(y_t;\mu_{y,t},\Omega_{y,t})=(2\pi)^{-d/2}\det(\Omega_{y,t})^{-1/2}\exp\left\lbrace -\frac{1}{2}(y_t - \mu_{y,t})'\Omega_{y,t}^{-1}(y_t - \mu_{y,t}) \right\rbrace .
\end{equation}
That is, the conditional distribution is the $d$-dimensional Gaussian distribution with mean $\mu_{y,t}$ and covariance matrix $\Omega_{y,t}$. 

The GSTVAR model has the advantage that, as the conditional distribution is Gaussian, the stationary distributions of the regimes corresponding to $p$ consecutive observations are known, and, hence, the weighted relative likelihoods of the regimes can be used as transition weights. In this specification, the transition weights depend on the full distribution of the preceding $p$ observations, and they are defined identically to the mixing weights of the GMVAR model \citep{Kalliovirta+Meitz+Saikkonen:2016}. Denoting $\boldsymbol{y}_{t-1}=(y_{t-1},...,y_{t-p})$, the transition weights are defined as 
\begin{equation}\label{eq:alpha_mt}
\alpha_{m,t} = \frac{\alpha_m n_{dp}(\boldsymbol{y}_{t-1};\boldsymbol{1}_p\otimes \mu_m, \boldsymbol{\Sigma}_{m,p})}{\sum_{n=1}^M \alpha_n n_{dp}(\boldsymbol{y}_{t-1};\boldsymbol{1}_p\otimes \mu_n, \boldsymbol{\Sigma}_{n,p})}, \ \ m=1,...,M,
\end{equation}
where $\alpha_1,...,\alpha_M$ are transition weight parameters that satisfy $\sum_{m=1}^M \alpha_m=1$ and $n_{dp}(\cdot;\boldsymbol{1}_p\otimes \mu_m, \boldsymbol{\Sigma}_{m,p})$ is the density function of the $dp$-dimensional normal distribution with mean $\boldsymbol{1}_p\otimes \mu_m$ and covariance matrix $\boldsymbol{\Sigma}_{m,p}$. The symbol $\boldsymbol{1}_p$ denotes a $p$-dimensional vector of ones, $\otimes$ is Kronecker product, $\mu_m=(I_d - \sum_{i=1}^pA_{m,i})^{-1}\phi_{m,0}$, and the covariance matrix $\boldsymbol{\Sigma}_{m,p}$ is given in \citet[Equation~(2.1.39)]{Lutkepohl:2005}, but using the parameters of the $m$th regime. In other words, $n_{dp}(\cdot;\boldsymbol{1}_p\otimes \mu_m, \boldsymbol{\Sigma}_{m,p})$ corresponds to the density function of the stationary distribution of the $m$th regime.

The transition weights are thus weighted ratios of the stationary densities of the regimes corresponding to the preceding $p$ observations. The specification of the transition weights is appealing, as it states that the greater the weighted relative likelihood of a regime is, the greater the weight of this regime is. The regimes are, hence, formed based on the statistical properties of the data and are not affected by the choice of  switching variables. This is a convenient feature for forecasting, and it also facilitates associating statistical characteristics and economic interpretations to the regimes.


Our GSTVAR model has a number of desirable features compared to popular alternative nonlinear VAR models. As already discussed, in line with the GMVAR model of \cite{Kalliovirta+Meitz+Saikkonen:2016}, the transition between the regimes depends on the full distribution of multiple past observations. By contrast, in the Markov-switching VAR (MS-VAR) model,the transition probabilities depend only on the most recent regime, whereas in the logistic STVAR (LSTVAR) and threshold VAR (TVAR) models, the regime is determined by the lagged value(s) of some observable variable(s) or exogenous switching variable(s). On the other hand, like the LSTVAR model, the GSTVAR model facilitates gradual shifts between the regimes, whereas in the TVAR, GMVAR and MS-VAR models each observation is generated from a single regime. Hence, the GSTVAR model combines several advantages of the GMVAR, TVAR, LSTVAR and MS-VAR models. A more detailed comparison of the models is given in Appendix~\ref{sec:comparison}.


\section{Estimation and model selection}\label{sec:estimation}

\subsection{Maximum likelihood estimation}

The parameters of the GSTVAR model can be estimated by the method of maximum likelihood (ML). We collect the parameters to the vector 
$
\boldsymbol{\theta}=(\phi_{1,0},...,\phi_{m,0},\varphi_1,...,\varphi_M,\sigma,\alpha)
$,
where $\varphi_m=(\text{vec}(A_{m,1}),....,\text{vec}(A_{m,p}))$, $m=1,...,M$, $\sigma=(\text{vech}(\Omega_1),...,\text{vech}(\Omega_M))$, and $\alpha=(\alpha_1,...,\alpha_{M-1})$ contains the transition weight parameters (notice that $\alpha_M=1-\sum_{m=1}^{M-1} \alpha_m$). 

Indexing the observed data as $y_{-p+1},...,y_0,y_1,...,y_T$, the conditional log-likelihood function, conditional on the initial values $\boldsymbol{y}_0=(y_0,...,y_{-p+1})$, is given as
\begin{equation}\label{eq:loglik1}
L_t(\boldsymbol{\theta})=\sum_{t=1}^T \log n_d(y_t;\mu_{y,t},\Omega_{y,t}).
\end{equation}
where $n_d(y_t;\mu_{y,t},\Omega_{y,t})$ is the $d$-dimensional conditional density of the process, conditional on $\mathcal{F}_{t-1}$, at time $t$, given in Equation~(\ref{eq:gausconddist}). Hence, the ML estimator of $\boldsymbol{\theta}$ maximizes the log-likelihood function $L_t(\boldsymbol{\theta})$ over the parameter space specified below.

To ensure ergodic stationarity of the process, we assume that Condition~\ref{cond:sufficient} in Section \ref{sec:genstvar} holds. Moreover, it is assumed that the true parameter value is an interior point of a compact subset of the parameter space, which is a standard condition for asymptotic normality of the ML estimator. Thus, given ergodic stationarity of the process, there is no particular reason to expect that the standard asymptotic results of consistency and asymptotic normality would not apply to the ML estimator. Finally, to achieve identification of the parameters such that the regimes cannot 'relabelled' to obtain the same model with different parameter vectors,we order the transition weight parameters $\alpha_m$, $m=1,...,M$ in a decreasing order. That is, we assume
\begin{equation}\label{eq:identcond}
\alpha_1>...>\alpha_M \text{ and } (\phi_{m,0},\varphi_m,\text{vech}(\Omega_m))\neq (\phi_{n,0},\varphi_n,\text{vech}(\Omega_n)) \text{ for } m\neq n\in\lbrace 1,...,M\rbrace .
\end{equation}
The constraints imposed on the parameter space are summarized in the following assumption.
\begin{assumption}\label{as:mle}
The true parameter value $\boldsymbol{\theta}_0$ is an interior point of $\boldsymbol{\Theta}$, which is a compact subset of the parameter space
$\lbrace \boldsymbol{\theta}=(\phi_{1,0},...,\phi_{m,0},\varphi_1,...,\varphi_M,\sigma,\alpha)\in\mathbb{R}^{M(d + d^2p + d(d+1)/2)}\times (0,1)^{M-1}:$ $\Omega_m$ is positive definite for all $m=1,...,M$, and Conditions~\ref{cond:sufficient} and (\ref{eq:identcond}) hold.$\rbrace$.
\end{assumption}
%

Finding the ML estimate amounts to maximizing the log-likelihood function in (\ref{eq:loglik1}) over a high dimensional parameter space satisfying the constraints in Assumption~\ref{as:mle}. Due to the complexity of the log-likelihood function, numerical optimization methods are required. The maximization problem can, however, be challenging in practice due to the complicated dependence of the transition weights on the preceding observations, which induces a large number of modes of the log-likelihood function and large areas of the parameter space, where it is flat in multiple directions. 

We follow \cite{Virolainen:2022, Virolainen:2025} (and others) and employ a two-phase estimation procedure that is run for a large number of times.\footnote{For example, in our empirical application presented in Section~\ref{sec:empapp}, we use $6000$ estimation rounds to estimate the four-variate GSTVAR with $p=4$, $M=2$.} In the first phase, a genetic algorithm is used to find parameter values that lie close to the local maximum points of the log-likelihood function. Since genetic algorithms tend to converge slowly near local solutions, a gradient based variable metric algorithm is run for each of the starting values, resulting in a number of alternative local solutions. Some of the estimation rounds may end up in saddle points or near-the-boundary points that are not local solutions, and some of the local solutions may be  inappropriate for statistical inference (for instance, there might be only few observations from some of the regimes). Because Condition~\ref{cond:sufficient} included in Assumption \ref{as:mle} is computationally costly to verify, we recommend using the necessary Condition \ref{cond:necessary} to restrict the parameter space.

After the estimation rounds have been run, the researcher can choose among the appropriate local solutions the one that maximizes the log-likelihood. Then, it can be verified that the selected local solution satisfies the sufficient Condition~\ref{cond:sufficient} for ergodic stationarity if the more easily checked necessary Condition~\ref{cond:necessary} is used to restrict the parameter space in estimation. The accompanying R package sstvars \citep{sstvars} employs a modified genetic algorithm that works similarly to that described in \cite{Virolainen:2022}.

To study the finite-sample properties of the ML estimator, we conduct a small scale Monte Carlo study, which is discussed in detail in Appendix~\ref{sec:montecarlo}. Because estimation is computationally demanding, we consider a simple bivariate GSTVAR model with $M=2$ and $p=1$. We consider two specifications that differ with respect to the values of the autoregressive parameters and intercepts (see Table~\ref{tab:monteparams} in Appendix~\ref{sec:montecarlo}). According to the results (see Table~\ref{tab:monteresults} in Appendix~\ref{sec:montecarlo}) the estimator is slightly biased in small samples, but the bias vanishes and estimation accuracy increases with the sample size. Also, estimation accuracy does not seem to depend on the specification. 

\subsection{Model selection}\label{sec:modelselection}

To select the number of regimes and the autoregressive order of the GSTVAR model, a suitable strategy is to start with a relatively simple specification and then build up to more complicated models if necessary. In particular, it may be useful to begin with linear Gaussian VAR models (i.e., GSTVAR models with only one regime) to evaluate to what extent they can capture the relevant characteristics of the data. Then, the order of the model or the number of regimes can be increased, if needed.

It is well known that testing for linearity against a model with multiple regimes, in general, poses a nonstandard testing problem because the model is identified only under the alternative \citep{Davies:1977}. This is the case also in our setup, and, therefore, we follow the previous literature (e.g., \citealp{Kalliovirta+Meitz+Saikkonen:2016} and \citealp{Virolainen:2022, Virolainen:2025}) and recommend using information criteria and residual-based diagnostic checks to compare the fit of models with different orders and numbers of regimes, as well as to study their adequacy in capturing the autocorrelation structure, conditional heteroskedasticity, and distribution of the data. The derivation of formal diagnostic tests is beyond the scope of this paper, but graphical devices, including quantile-quantile plots as well as autocorrelation and cross-correlation functions of the residuals and their squares, can be used to this end. 

For a number of reasons, GSTVAR models of a relatively low autoregressive order are preferable. Firstly, the estimation of GSTVAR models with a large $p$ can be tedious in practice. Secondly, if the order $p$ is large, the number of parameters increases vastly when the number of regimes is increased, which is likely to substantially decrease estimation accuracy. Thirdly, unlike linear VARs, decreasing the autoregressive order may actually improve the fit because the transition weights are calculated using the whole joint distribution of the preceding $p$ observations. Consequently, the sensitivity of the transition weights to individual observations and, hence, responsiveness to changes in the dynamics of the data decrease with $p$.

It is also advisable to be conservative with the number of regimes $M$ because if the number of regimes is chosen too large, some of the parameters in the model are not identified \citep[see the related discussion in][Section~3.4]{Kalliovirta+Meitz+Saikkonen:2016}. Having too many regimes may also lead to overfitting, as the number of parameters may become overly large compared to the number of observations in some of the regimes. Moreover, increasing the number of regimes substantially increases the complexity of the surface of the log-likelihood function, making estimation of the parameters challenging in practice.

\section{Empirical application}\label{sec:empapp}

In this section, we apply our GSTVAR model to study the effects of severe weather on the U.S. macroeconomy. This issue was recently addressed by  \cite{Kim+Matthes+Phan:2022}, who employ a two-regime recursive STVAR model with transition weights defined by a linear time trend ($\alpha_{2,t}=t/T$ and $\alpha_{1,t}=1-\alpha_{2,t}$ in \eqref{eq:stvar1}). They argue that allowing for time-varying coefficients in the model is important because the strength of the economic impact of severe weather may have diminished due to actions potentially taken to make the economy more adaptable to climate change. However, they find little evidence in favor of adaptation to severe weather in the U.S, but on the contrary, their results suggest that the severe weather shock has a significant effect only towards the end of the sample (spanning from 1963 to 2019).

Using transition weights defined by a linear time trend in a two-regime STVAR model only allows one smooth transition from one regime to the other during the sample period. This approach may be too simple to describe the adaptation of the economy, as the effects of severe weather may vary across different states of the economy due to variation in factors such as fiscal flexibility and sector-specific vulnerabilities. However, it is not obvious which exogenous or lagged endogenous variables would be capable of capturing such variation in the transition weights in a nonlinear SVAR model. Therefore, we address this issue with our GSTVAR model in which the regimes and shifts between them are formed based on the statistical properties of the data through the full distribution of the preceding $p$ observations. Due its more data-driven nature, this approach plausibly better describes the evolution of the joint dynamics of severe weather and U.S. macroeconomy through time than an STVAR model with transition weights defined by a linear time trend or depending only on the level of exogenous or lagged endogenous variables. 



\subsection{Data}

We consider a four-variable monthly U.S. data set that comprises indicators of severe weather and economic activity as well as consumer price inflation and an interest rate variable. While our sample period (from 1961:1 to 2022:3, $T=735$ observations) is somewhat longer than than in \cite{Kim+Matthes+Phan:2022}, the main conclusions based on their shorter sample period (from 1963:4 to 2019:5) remain the same (see Appendix \ref{sec:rob_subsamp} for the subsample results). The end point of our sample period is determined by the availability of the monthly GDP growth data, but as argued by \cite{Kim+Matthes+Phan:2022}, it is important to use data of higher frequency than quarterly because some weather effects can be short-lived. 

As an indicator of the frequency of severe weather and the extent of sea level rise, we use the Actuaries Climate Index (ACI), developed by actuarial organizations in the United States and Canada \citep{ACI:2023}. Following \cite{Kim+Matthes+Phan:2022}, we seasonally adjust the ACI series with the standard Census Bureau X-13 seasonal adjustment algorithm. For a more detailed description of ACI, see \cite{Kim+Matthes+Phan:2022}. 

As an aggregate measure of real economic activity, we use the monthly GDP growth rate constructed by the Federal Reserve Bank of Chicago from a collapsed dynamic factor analysis of a panel of $500$ monthly measures of real economic activity and quarterly real GDP growth \citep{MGDP:2023}. While \cite{Kim+Matthes+Phan:2022} use both the industrial production index and unemployment rate to measure real economic activity at the monthly frequency, we opt for the monthly GDP growth rate instead. It has the advantage of being a more comprehensive measure of real economic activity, including agricultural output, tourism, and services, which can be significantly affected by the severe weather shocks. Moreover, using only a single variable to measure real economic activity is likely to facilitate estimation by reducing the dimension, and thereby also the number of parameters, of the model.

\begin{figure}[t]
    \centerline{\includegraphics[width=\textwidth - 2cm]{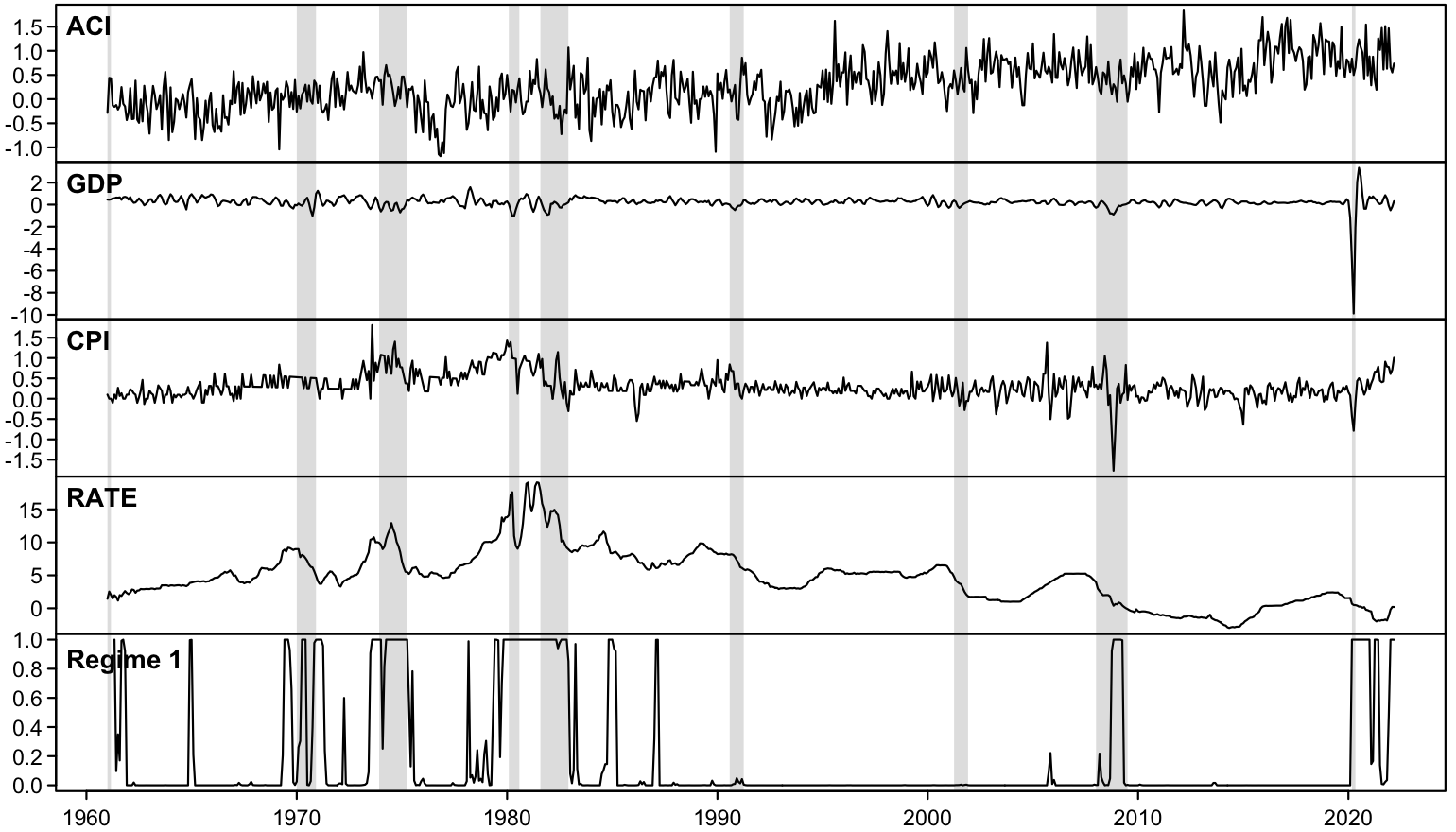}}
    \caption{Monthly U.S. time series covering the period from 1961:1 to 2022:3. From top to bottom, the variables are the Actuaries Climate Index, the monthly GDP growth rate, the monthly CPI growth rate, and the effective federal funds rate (replaced by the \cite{Wu+Xia:2016} shadow rate for the zero lower bound periods). The bottom panel shows the estimated transition weights of the first regime of the fitted two-regime fourth-order GSTVAR model. The shaded areas indicate the U.S. recessions defined by the NBER.}
\label{fig:seriesplot}
\end{figure}

Finally, we include the monthly growth rate of the consumer price index (CPI) and the effective federal funds rate (RATE) that is replaced by the \cite{Wu+Xia:2016} shadow rate for the zero lower bound periods. 
The CPI and federal funds rate series were downloaded from the Federal Reserve Bank of S.t. Louis database and \cite{Wu+Xia:2016} shadow rate from the Federal Reserve Bank of Atlanta database. The time series are depicted in the top four panels of Figure~\ref{fig:seriesplot}, where the shaded areas indicate the U.S. recessions defined by the NBER. 

\subsection{GSTVAR model}

To select the order of the GSTVAR model, we start by examining the partial autocorrelation functions of the U.S. time series (shown in Figure~\ref{fig:pacf} of Appendix~\ref{sec:detailsempapp}). They suggest that the autoregressive order $p=4$ might be adequate. Therefore, we fit two-regime GSTVAR models with autoregressive orders $p=1,....,5$ and, to compare them, compute the values of three information criteria (AIC, BIC and HQIC). As seen in Table \ref{tab:ic}, the order $p=4$ minimizes the AIC, whereas the order $p=2$ minimizes the HQIC and BIC. The table also contains similar results for linear VAR (i.e., GSTVAR with $M=1$) models, which are clearly inferior to the two-regime GSTVAR models in terms of the information criteria. Moreover, the constancy of the AR matrices and intercepts as well as the constancy of AR matrices only are clearly rejected by the Wald test (see Appendix~\ref{sec:detailsempapp}). While adding a third regime to the model might improve the fit, the number of parameters compared the number of observations in each regime would increase substantially, possibly leading to overfitting. Moreover, since incorporating too many regimes in the model could also result in identification issues (see Section~\ref{sec:modelselection}), we confine ourselves to two-regime models.

\begin{table}
\centering
\small
\renewcommand{\arraystretch}{1.0}
\begin{tabular}{c c c c c c}
Model      & Log-lik  & BIC     & HQIC    & AIC  \\ 
\hline\\[-2.5ex]
$p=1,M=1$  & $-1.755$ & $3.780$ & $3.665$ & $3.593$ \\
$p=2,M=1$  & $-1.460$ & $3.334$ & $3.157$ & $3.046$ \\
$p=3,M=1$  & $-1.411$ & $3.381$ & $3.142$ & $2.992$ \\
$p=4,M=1$  & $-1.376$ & $3.456$ & $3.155$ & $2.966$ \\
$p=5,M=1$  & $-1.355$ & $3.560$ & $3.197$ & $2.968$ \\
$p=1,M=2$  & $-0.548$ & $1.645$ & $1.410$ & $1.263$ \\
$p=2,M=2$  & $-0.220$ & $1.277$ & $0.919$ & $0.694$ \\
$p=3,M=2$  & $-0.181$ & $1.489$ & $1.007$ & $0.704$ \\
$p=4,M=2$  & $-0.121$ & $1.658$ & $1.052$ & $0.671$ \\
$p=5,M=2$  & $-0.130$ & $1.968$ & $1.237$ & $0.778$
\end{tabular}
\caption{The values of the log-likelihood function and the information criteria (divided by the number of observations) for a number of linear VAR and two-regime GSTVAR models.}
\label{tab:ic}
\end{table}


Based on graphical residual diagnostics, presented in Appendix~\ref{sec:detailsempapp}, the two-regime GSTVAR model with $p=4$ lags captures the autocorrelation structure of the data reasonably well. Conditional heteroskedasticity and marginal distribution of the series are not completely captured, but the inadequacies are not particularly severe. 
Finally, the sufficient Condition~\ref{cond:sufficient} for ergodic stationarity holds for the selected model, as the upper bound of the joint spectral radius of the matrices $\boldsymbol{A}_m$ (\ref{eq:boldA}), $m=1,2$ (see Section \ref{sec:genstvar}), is found to be strictly less than one ($0.995$). Hence, we proceed with the two-regime fourth-order model.

The estimated transition weights indicate the relative importance of each regime. The time series of the weights of Regime~1 are presented in the bottom panel of Figure~\ref{fig:seriesplot}, and the corresponding weights of Regime~2 are, of course, obtained by subtracting these weights from unity. The shifts between the regimes are relatively fast, with the switch from one regime to the other typically taking from one to three months to be completed. Regime~1 mainly dominates in the 1960s, 1970s, and 1980s, but obtains large weights also during the Financial crisis and from the beginning of the COVID-19 crisis onward (excluding a short period in 2021). Overall, Regime~1 thereby mostly prevails in the earlier sample and Regime~2 in the later sample. The stationary standard deviations of ACI, GDP growth rate, CPI growth rate, and interest rate in Regime~1 are $0.59, 1.17, 0.47$, and $9.57$, respectively, whereas they are $0.52, 0.29, 0.26$, and $3.46$ in Regime~2. These differences and the dominance of Regime~1 in the volatile periods of 1970s and 1980s as well as during the Financial crisis and the COVID-19 crisis, suggest that it could represent more turbulent times compared to Regime~2.

\subsection{Structural analysis}\label{sec:struct_anal}

We are predominantly interested in the economic impact of severe weather, and, therefore, a severe weather shock  (ACI shock) must be identified. This amounts to imposing a unique structure of the impact matrix $B_t$ in (\ref{eq:stvar3}) governing the contemporaneous relationships of the shocks, so that one of the shocks can be labelled the ACI shock. Following \cite{Kim+Matthes+Phan:2022}, our key identification restriction is that the ACI shock is the only shock that can instantaneously affect the ACI. This identification restriction seems reasonable, as it states that the other (macroeconomic) shocks do not affect severe weather or sea level within a month, but allows them to affect the ACI in the long run. We establish the identification by placing the ACI first in the vector of variables (with the rest of the variables in the order GDP, CPI, RATE) and imposing a recursive lower-triangular structure on the impact matrix $B_t$, obtained by the Cholesky decomposition of the estimated conditional error covariance matrix $\Omega_{y,t}$.

To study the effects of the ACI shock, we compute the generalized impulse response functions of the variables to it. As pointed out in Section~\ref{sec:SSTVAR}, since the transition weights are endogenously determined and the regime can shift as a result of a shock, the impulse responses generally depend on the initial values as well as on the sign and size of the shock. We are particularly interested in finding out about the potential state-dependence of the effects of the ACI shock. To capture them, we take all such histories of length $p$ from the data that indicate the dominance of a given regime, and compute the GIRFs for each regime separately. Specifically, for Regime~$\tilde{m}$, we take the histories $\boldsymbol{y}_{t-1}$ for which the corresponding transition weight $\alpha_{\tilde{m},t}$ is greater than $0.75$. With the threshold $0.75$, the given regime is clearly dominant but a reasonably large amount of histories are still included in both regimes (Regimes~1 and 2 have $115$ and $603$ such histories, respectively). To closely match the properties of the data, for each history, the GIRF to the corresponding recovered structural shock is computed. Finally, the GIRFs are scaled so that the instantaneous response of the ACI is $0.3$, making the responses comparable.


\begin{figure}[!t]
    \centerline{\includegraphics[width=\textwidth - 2cm]{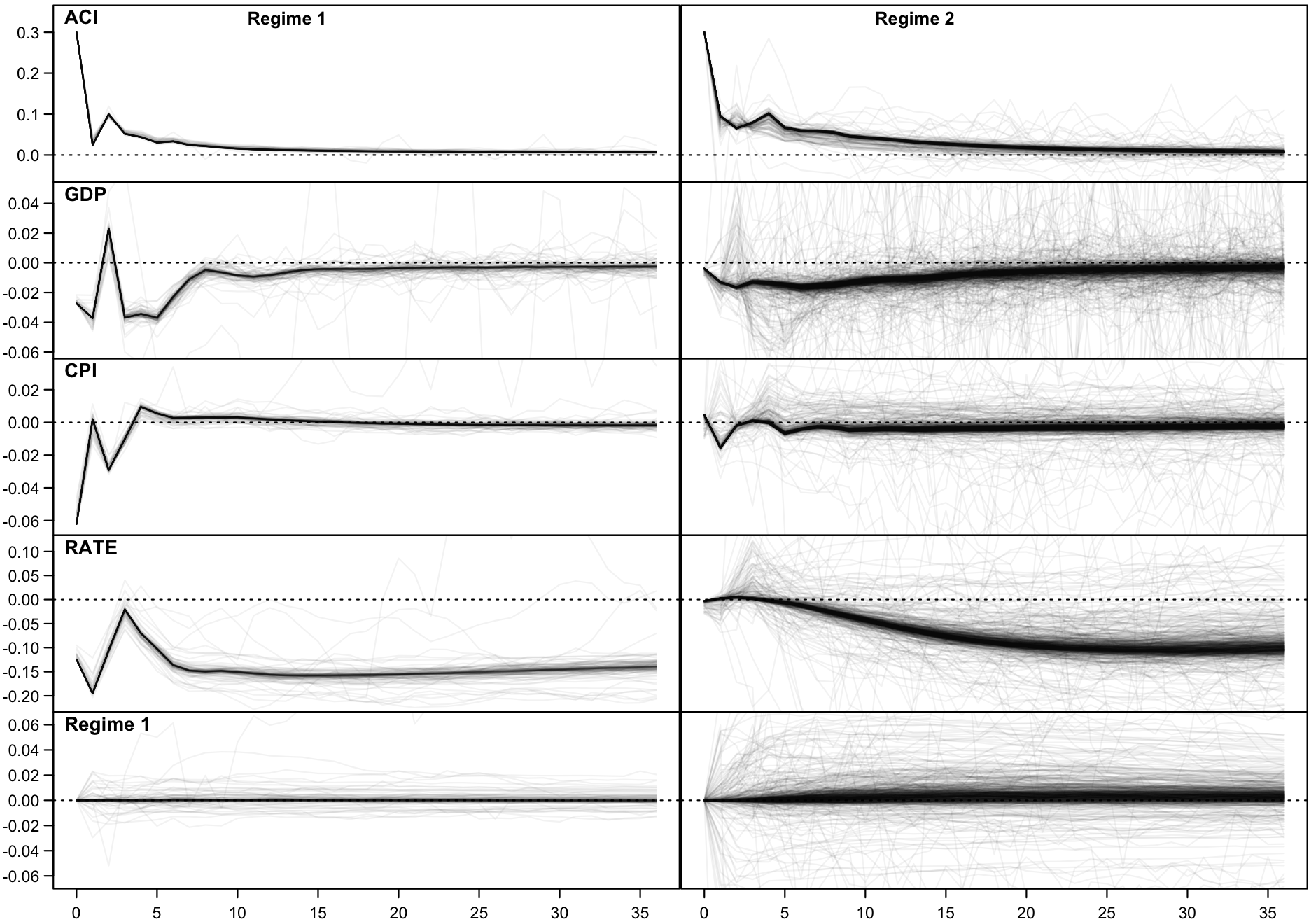}}
    \caption{Generalized impulse response functions to the identified ACI shock $h=0,1,....,36$ months ahead. From top to bottom, the responses of ACI, GDP growth rate, CPI growth rate, the interest rate variable, and transition weights of Regime 1 are depicted in the rows. The left and right columns show the responses conditional on Regime~1 and Regime~2 dominating when the shock arrives, respectively. All GIRFs have been scaled so that the instantaneous change in the ACI index is $0.3$. The results are based on $2500$ Monte Carlo repetitions for each history (for details of the algorithm, see Appendix~\ref{sec:montecarlo_girf}).}
\label{fig:girfplot}
\end{figure}

Figure~\ref{fig:girfplot} presents the sample distribution (over the histories) of the GIRFs of the variables, including the transition weights, to the identified ACI shock $h=0,1,...,36$ quarters ahead. Each light gray curve with low opacity depicts the GIRF corresponding to one history $\boldsymbol{y}_{t-1}$, so the darkness of a region in the figure indicates a greater concentration of GIRFs in that area. The left and right columns depict the responses in Regime~1 and Regime~2, respectively. In Regime~1, the vast majority of the GIRFs follow virtually the same sample paths. This is the case because the ACI shock has almost no effect on the transition weights of Regime~1, as shown in the bottom left panel. Although the majority of the GIRFs follow similar sample paths also in Regime~2, they exhibit more variation depending on the history $\boldsymbol{y}_{t-1}$ (and the related sign and size of the shock). 

In both regimes, a positive ACI shock mainly decreases GDP, consumer prices, and the interest rate. However, there is a major difference in the GIRFs between the regimes: the effects of the ACI shock on all macroeconomic variables are clearly stronger in Regime~1. This may be the case because Regime~1 appears to be attributed to more volatile times, including recessions and crisis periods. Thus, it might represent a state of the economy that is more vulnerable to weather related shocks, and therefore, a similar shock is likely to result in greater damage in Regime~1 than in Regime~2. Other differences between the regimes are minor. 


As Regime~1 dominates the earlier part and Regime~2 is predominant in the latter part of the sample period, it seems reasonable to conclude that the effect of severe weather shocks has diminished over time. This suggests that the U.S. economy has adapted to the effects of climate change. In contrast, \cite{Kim+Matthes+Phan:2022} found the effects of the ACI shock stronger at the end than at the beginning of their sample period from 1963:4 to 2019:5, which they interpreted as insufficient adaptation of the U.S. economy to the changing distribution of weather related shocks. 
While our results are thus, in general, contradictory to theirs, there are some short periods in the latter part of our sample where Regime~1 with stronger impact of the ACI shock prevails. These particularly overlap with the major crises, which suggests that the potential adaptation may not provide sufficient resilience against weather related shocks in a turbulent state of the economy. 

\begin{table}[t]
\centering
\small
\renewcommand{\arraystretch}{1.0}
\begin{tabular}{c l@{\hspace{1cm}}cccccc}
& & $h=0$ & $h=3$ & $h=6$ & $h=12$ & $h=24$ & $h=36$ \\ 
\hline
\noalign{\smallskip}
\textbf{ACI} & All data  & $100.00\%$ & $85.73\%$ & $80.32\%$ & $77.22\%$ & $72.69\%$ & $69.73\%$ \\
             & Regime 1  & $100.00\%$ & $73.04\%$ & $67.88\%$ & $62.06\%$ & $55.24\%$ & $51.50\%$ \\
             & Regime 2  & $100.00\%$ & $88.25\%$ & $82.81\%$ & $80.39\%$ & $76.44\%$ & $73.44\%$ \\
             
\hline
\noalign{\smallskip}
\textbf{GDP} & All data  & $5.38\%$ & $10.60\%$ & $14.34\%$  & $16.74\%$ & $17.57\%$ & $17.39\%$ \\
             & Regime 1  & $5.96\%$ & $\phantom{1}4.49\%$  & $\phantom{1}6.37\%$  & $\phantom{1}6.12\%$  & $\phantom{1}5.93\%$  & $\phantom{1}5.77\%$ \\
             & Regime 2  & $5.25\%$ & $11.83\%$ & $16.01\%$ & $18.97\%$ & $20.03\%$ & $19.85\%$ \\
\hline
\noalign{\smallskip}
\textbf{CPI} & All data  & $\phantom{1}5.84\%$  & $\phantom{1}7.70\%$  & $7.45\%$  & $7.25\%$ & $7.43\%$ & $7.47\%$ \\
             & Regime 1  & $18.92\%$ & $11.01\%$ & $9.93\%$  & $8.94\%$ & $8.19\%$ & $7.86\%$ \\
             & Regime 2  & $\phantom{1}3.39\%$  & $\phantom{1}7.17\%$  & $7.07\%$  & $7.02\%$ & $7.36\%$ & $7.50\%$ \\
\hline
\noalign{\smallskip}
\textbf{RATE} & All data  & $\phantom{1}3.53\%$  & $2.21\%$ & $2.21\%$ & $5.72\%$  & $11.88\%$ & $14.58\%$ \\
              & Regime 1  & $15.30\%$ & $9.66\%$ & $8.66\%$ & $10.23\%$ & $11.90\%$ & $12.30\%$ \\
              & Regime 2  & $\phantom{1}1.14\%$  & $0.73\%$ & $0.97\%$ & $4.92\%$  & $12.05\%$ & $15.21\%$ \\
\hline
\noalign{\smallskip}
$\boldsymbol{\alpha_{1,t}}$ & All data  & & $13.53\%$ & $13.77\%$ & $14.66\%$ & $14.71\%$ & $13.72\%$ \\
                            & Regime 1  & & $\phantom{1}5.55\%$  & $\phantom{1}5.20\%$  & $\phantom{1}4.51\%$  & $\phantom{1}4.49\%$  & $\phantom{1}4.71\%$ \\
                            & Regime 2  & & $15.24\%$ & $15.60\%$ & $16.81\%$ & $16.85\%$ & $15.62\%$ \\
\end{tabular}
\caption{The relative contribution of the ACI shock to the forecast error variances of the variables in each regime at the horizons $h=0,3,6,12,24,36$ based on the generalized forecast error variance decompositions calculated using the length $p$ histories of the data with the related sizes and signs of the shock. The rows labeled "All data" present the GFEVDs based on GIRFs using all the length $p$ histories of the data with the relates signs and sizes of the shock, whereas rows labeled "Regime~1" or "Regime~2" shows the GFEVDs based on only such length $p$ histories that indicate the dominance of the specified regime (i.e., the its related transition weight is larger than $0.75$). The relative contribution to the transition weights $\alpha_{1,t}$ is not defined (and thereby not shown) at impact, because the transition weights are $\mathcal{F}_{t-1}$-measurable.}
\label{tab:gfevd}
\end{table}

To assess the relative importance (or economic significance) of the ACI shock, we compute the generalized forecast error variance decompositions using all the histories $\boldsymbol{y}_{t-1}$, $t=1,...,T$, in the data as the initial values with the corresponding sign and size of the recovered structural shock. The rows labeled ``All data" in Table~\ref{tab:gfevd} present the relative contribution of the ACI shock to the forecast error variance of the variable in question at each horizon, while the rows labeled ``Regime~1" and ``Regime~2" contain the corresponding figures for each regime (i.e., for the histories with a transition weight of the regime in question higher than $0.75$ in time period $t$). The ACI shock dominates the ACI index, but the other shocks become relatively more important at longer horizons. For the remaining variables, the ACI shock plays a lesser, yet clearly notable, role, contributing approximately 5--17\%, 6--8\% and 2--15\% to the forecast error variance of GDP, CPI and the interest rate in the entire sample period, respectively. The ACI shock also seems to have some importance for regime shifts, with its relative contribution to the forecast error variance of the transition weight $\alpha_{1,t}$ hovering between 13 and 15\%. The ACI shock is economically significant also in both regimes. Compared to Regime~1, its relative importance is, in general, greater for GDP and the transition weight in Regime~2, whereas the opposite holds for CPI and the interest rate, especially at short horizons.

\subsection{Robustness checks}
To check the robustness of our findings with respect to the sample period, we also consider the subsample period from 1963:4 to 2019:5 considered by \cite{Kim+Matthes+Phan:2022}. The results general conclusion remain intact in that the regimes of the fitted GSTVAR model quite similar to our main specification. Also, the estimated GIRFs and are similar, with the exception that the interest rate variable increases in response to a positive ACI shock in Regime 1. The details, including the evolution of the transition weights and the GIRFs can be found in Appendix~\ref{sec:rob_subsamp}.

For comparison, we also report the results of the GMVAR model \citep{Kalliovirta+Meitz+Saikkonen:2016} as well as the recursively identified linear Gaussian SVAR model (obtained as special case of GMVAR and GSTVAR models with $M=1$) for the entire sample period. The mixing weights of the fitted GMVAR model, presented in Figure~\ref{fig:rob_weightplot} in Appendix~\ref{sec:rob_gmvar} are rather similar to the transition weights of our GSTVAR model, as expected due to the similar function forms of the weight functions. Also the GIRFs, presented in Figure~\ref{fig:rob_girfplot} in Appendix~\ref{sec:rob_gmvar}, are quite similar to those in our GSTVAR model. In particular, in line with our GSTVAR model, the GMVAR model displays clearly stronger responses of GDP, inflation and the interest rate to the ACI shock in the more turbulent Regime~1 than in Regime~2. 
Finally, the finding that a positive ACI shock causes a decrease in all GDP, inflation and the interest is quite robust, as it emerges in all models considered, including the recursive Gaussian SVAR model, whose impulse responses are also depicted in Figure~\ref{fig:rob_girfplot} in Appendix~\ref{sec:rob_gmvar}.

\section{Conclusion}\label{sec:conclusion}
We have introduced a Gaussian smooth transition vector autoregressive (GSTVAR) model with transition weights adopted from the Gaussian mixture VAR (GMVAR) model of \cite{Kalliovirta+Meitz+Saikkonen:2016}. However, compared to the GMVAR model, our model has the advantage of great flexibility in that it enables capturing gradual shifts in the dynamics of the data, whereas the GMVAR model involves discrete regimes. On the other hand, switching between the regimes in the $pt$h-order GSTVAR model depends on the full distribution of the preceding $p$ observations, in contrast to previously introduced other smooth transition VAR models, in which it is governed by specific switching variables. Hence, our model facilitates associating the regimes with statistical information in the data in a versatile manner. The structural counterpart of the GSTVAR model has the desirable feature that it produces the dynamic effects of the structural shocks on the transition weights governing regime switching. 

We have discussed estimation, model selection and diagnostic checking as well as conducting structural analysis in the GSTVAR model. Moreover, by making use of the results of \cite{Saikkonen:2008} and \cite{Kheifets+Saikkonen:2020}, we have introduced a sufficient condition for stationarity and ergodicity of the model. Due to the complexity of the log-likelihood function, estimation by the method of maximum likelihood can be quite tedious in practice, and following \cite{Virolainen:2022, Virolainen:2025} (and others), we propose a two-phase procedure in which a genetic algorithm is used to find starting values for a gradient based estimation method. The introduced methods, including a modified version of a genetic algorithm, are implemented in the CRAN distributed R package sstvars \citep{sstvars}. A potential area of future research is an extension of our model that accommodates conditional heteroskedasticity by incorporating autoregressive conditional heteroskesticity or stochastic volatility to the shocks. 

We have applied our method to study the macroeconomic effects of severe weather shocks in the U.S. Our monthly data set covers the period from 1961:1 to 2022:3 and contains an indicator of the frequency of severe weather and the extent of sea level rise in addition to a number of macroeconomic variables. Our structural GSTVAR model has two regimes, one of which (Regime~2) prevails in the latter part of the sample period, while the other regime (Regime~1) mainly dominates its earlier part, particularly the turbulent times of 1970s and 1980s, but also prevails in the later sample during the Financial crisis and the COVID-19 crisis. \cite{Kim+Matthes+Phan:2022} have recently also studied the economic impact of severe weather in the U.S., and following their lead, we have identified the severe weather shock recursively, placing the severe weather indicator first in the vector of variables. A positive weather shock was found to decrease GDP, consumer prices, and the interest rate in both regimes, but the effects are stronger in Regime~1. Hence, our results suggest that the U.S. economy, with the exception of certain crisis periods, has adapted to the changing distribution of weather related shocks. This finding contradicts the main result of \cite{Kim+Matthes+Phan:2022}, who found the impact of the weather shock to get stronger over time, interpreting this as lack of adaptation to changing weather.

\bibliography{masterrefs.bib}

\newpage
\begin{appendices}
\renewcommand{\thefigure}{\thesection.\arabic{figure}}
\renewcommand{\thetable}{\thesection.\arabic{table}}
\setcounter{figure}{0}    
\setcounter{table}{0}

\section{Sufficient condition for stationarity and ergodicity}\label{sec:statcond}
The sufficient condition for stationarity and ergodicity of our STVAR model, Condition~\ref{cond:sufficient}, differs slightly from to the one in \citet[][Assumption~R]{Kheifets+Saikkonen:2020}. This is because our parametrization of the STVAR model differs from \cite{Kheifets+Saikkonen:2020}. Their result is, nonetheless, applicable with our parametrization too, as is explained below. 

The ergodic stationarity condition of \cite{Kheifets+Saikkonen:2020} is based on the results of \cite{Saikkonen:2008}, who derived a comparable condition for smooth transition vector error correction models. Specifically, in the proof of their Theorem~1, \cite{Kheifets+Saikkonen:2020} explain how their specification can be obtained as a special from the more general nonlinear error correction model in \citet[Equation~(17)]{Saikkonen:2008}. Our model, described in Section~\ref{sec:genstvar}, is obtained as a special case from the model of \cite{Saikkonen:2008} in a similar fashion but slightly differently due to differences in the parametrization. In particular, for the first term on the right side of \cite{Saikkonen:2008}, Equation~(17), we consider the left side of \citet{Kheifets+Saikkonen:2020}, Equation~(6), with $h_s=\alpha_{m,t}$ and $\bar{B}_{sj}=A_{m,i}$. Similarly, for the third term of \cite{Saikkonen:2008}, Equation~(17), we consider the left side of \cite{Kheifets+Saikkonen:2020}, Equation~(7). Consequently, Condition~\ref{cond:sufficient} replaces Assumption~R of \cite{Kheifets+Saikkonen:2020} as the sufficient condition for stationarity and ergodicity.

\section{Monte Carlo algorithm for estimating the GIRF}\label{sec:montecarlo_girf}
We present a Monte Carlo algorithm for estimating the generalized impulse response function defined in Equation~(\ref{eq:girf}) for initial values $\boldsymbol{y}_{t-1}=(y_{t-1},...,y_{t-p})$. Our algorithm is adapted from \citet[pp. 135-136]{Koop+Pesaran+Potter:1996} and \citet[pp. 601-602]{Kilian+Lutkepohl:2017}. We assume that one is either interested on a known set $S$ of the histories $\boldsymbol{y}_{t-1}$ (e.g., certain histories or a history taken from the data), or the history $\boldsymbol{y}_{t-1}$ follows a known distribution $G$ (e.g., the stationary distribution of a specific regime). In the following, $y_{t+h}^{(i)}(\delta_j,\boldsymbol{y}_{t-1})$ denotes a realization of the process at time $t+h$ conditional on the structural shock of sign and size $\delta_j$ in the $j$th element of $e_t$ hitting the system at time $t$ and on the $p$ observations $\boldsymbol{y}_{t-1}=(y_{t-1},...,y_{t-p})$ preceding the time $t$, whereas $y_{t+h}^{(i)}(\boldsymbol{y}_{t-1})$ denotes an alternative realization conditional on the history $\boldsymbol{y}_{t-1}$ only.

For a single history $\boldsymbol{y}_{t-1}$ and given sign and size $\delta_j$ of the $j$th structural shock (that is of interest), the GIRF can be estimated with the following steps. 
\begin{enumerate}\addtocounter{enumi}{-1}
\item Set the horizon $H$ and the number of repetitions $R_1$.


\item Draw $H+1$ independent realizations of the structural shock $e_t$ from the $d$-dimensional standard normal distribution. Then, impose the sign and size $\delta_j$ in the $j$th element of the first drawn structural shock $e_t$ to obtain $e_t^*$. Finally, calculate the modified reduced form shock $u_t^*=B_te_t^*$.\label{step2}

\item Use the modified reduced form shock $u_t^*$ and the rest $H$ structural shocks $\varepsilon_t$ obtained from Step~\ref{step2} to compute realizations $y_{t+h}^{(i)}(\delta_j,\boldsymbol{y}_{t-1})$ for $n=0,1,...,H$, iterating forward. At $h=0$, the modified reduced form shock $u_t^*$ calculated from the structural shock $e_t^*$ in Step~\ref{step2} is used. From $h=1$ onwards, the $h+1$th structural shock $e_t$ is used to calculate the reduced form shock $u_{t+h}=B_{t+h}e_{t+h}$.

\item  Use the rest $H+1$ the structural shocks $e_t$ obtained from Step~\ref{step2} to compute realizations $y_{t+h}^{(i)}(\boldsymbol{y}_{t-1})$ for $h=0,1,...,H$, so that the non-modified reduced form shock $u_t=B_te_t$ is used to compute the time $h=0$ realization. Otherwise proceed similarly to the previous step.

\item Calculate $y_{t+h}^{(i)}(\delta_j,\boldsymbol{y}_{t-1}) - y_{t+h}^{(i)}(\boldsymbol{y}_{t-1})$.\label{step5}

\item Repeat Steps~\ref{step2}--\ref{step5} $R_1$ times and calculate the sample mean of $y_{t+h}^{(i)}(\delta_j,\boldsymbol{y}_{t-1}) - y_{t+n}^{(i)}(\boldsymbol{y}_{t-1})$ for $h=0,1,...,H$ to obtain an estimate of the GIRF$(h,\delta_j,\boldsymbol{y}_{t-1})$.\label{step6}

\end{enumerate}

To estimate the GIRF for a number of histories (and possibly signs and sizes of the shock), repeat the above steps for each of the initial values (and signs and sizes of the shocks), either taking them from the set $S$ or generating them from the distribution $G$. The sign and size $\delta_j$ of the shock can vary depending on the initial value or it can be fixed throughout. To obtain summary statistics on the distribution of the GIRFs over the initial values (or signs or sizes of the shock), the sample mean or some sample quantiles over the GIRFs can be calculated.

\section{Comparison to related models}\label{sec:comparison}

In this appendix, we compare the GSTVAR model to a number of alternative closely related models popular in the literature, namely the GMVAR, LSTVAR, TVAR, MS-VAR, and TVP-VAR models.

Although the transition weight function of the GSTVAR model is similar to the mixing weights of the GMVAR model, it differs from that model in other ways. Specifically, the GMVAR model generates each observation from a randomly selected distinct regime, while the GSTVAR model allows for smooth transitions between the regimes and thereby facilitates gradual shifts in the dynamics. This is the case because in the GMVAR model, the transition weights $\alpha_{m,t}$ are replaced by unobservable regime variables $s_{m,t}$ in Equations~(\ref{eq:stvar1})--(\ref{eq:stvar3}). These regime variables $s_{m,t}$, $m=1,...,M$, are such that at each $t$, exactly one of them takes the value one, whereas the rest take the value zero according to the probabilities $P(s_{m,t}=1|\mathcal{F}_{t-1})=\alpha_{m,t}$. Because the regime that generates each observation is thus unknown in the GMVAR model, the (structural) errors cannot, in general, be recovered from the data, while they are readily available in the GSTVAR model. This is particularly useful in structural analysis because it enables to compute the historical decomposition and conducting counterfactual analyses, and the recovered structural shocks also facilitate computing GIRFs that reflect the properties of the data more precisely (see Sections~\ref{sec:SSTVAR} and \ref{sec:empapp}). 

The LSTVAR model is a particular STVAR model with logistic transition weights, originally introduced to VAR models in \cite{Anderson+Vahid:1998}. A simple two-regime ($M=2$) logistic STVAR model commonly considered in the previous literature is obtained by defining the transition weights as
\begin{equation}\label{eq:logisticweights}
\alpha_{1,t}=1-\alpha_{2,t}, \ \ \alpha_{2,t}=[1 + \exp\lbrace -\gamma (y_{i,t-j}-c)\rbrace]^{-1},
\end{equation}
where $y_{i,t-j}$ is the $j$th lagged observation $(j\in\lbrace 1,....,p\rbrace)$ of the $i$th variable $(i\in\lbrace 1,....,d\rbrace)$, $c\in\mathbb{R}$ is a location parameter, and $\gamma > 0$ is a scale parameter. The location parameter $c$ determines the mid-point of the transition function, i.e., the value of the switching variable when the weights are equal. The scale parameter $\gamma$, in turn, determines the smoothness of the transitions (smaller $\gamma$ implies smoother transitions), and it is assumed strictly positive so that $\alpha_{2,t}$ is increasing in $y_{it-j}$. This simple setup can be generalized in various directions \citep[see, e.g.,][]{Terasvirta+Yang:2014}. It is also possible to use exogenous or deterministic switching variables instead of a lagged endogenous variable in the logistic transition weights~(\ref{eq:logisticweights}). However, in these cases, it may be difficult to determine whether the model is stationary, and potentially important dynamics are ignored, as the shocks are not allowed affect the transition weights, contradicting the premise that economic agents may alter their behavior when facing changing economic conditions. 

The logistic transition weights have the advantage that the regimes typically have clear interpretations and the switching variable can be defined to address some specific aspect of the empirical problem. For example, when using the lagged GDP growth as the switching variable, the regimes can be interpreted as recession and expansion. In comparison, the transition weights of the GSTVAR model depend on the full distribution of the preceding $p$ observations and can thereby capture more complicated switching-dynamics, which facilitates associating the regimes with various aspects of the phenomenon being modelled. As the transition weights of the GSTVAR model are not affected by the choice of switching variables but determined based on the weighted relative likelihoods of the regimes, they are more clearly data-driven and their own evolution is interesting as such. Nonetheless, since the resulting regimes are statistically determined, it is not always clear how to interpret them in terms of a given application. 


The threshold VAR (TVAR) model \citep{Tsay:1998} is otherwise similar to the LSTVAR model, but the regime switches are discrete. It is obtained as a special case of the STVAR model of Section~\ref{sec:genstvar} by assuming
\begin{equation}
\alpha_{m,t}=I(c_{m-1}<s_t<c_m), \ \ m=1,...,M,
\end{equation}
in Equation~(\ref{eq:stvar1}), where $I(\cdot)$ is an indicator function, $s_t$ is the switching variable at the time $t$, and $-\infty<c_{1}<\cdots <c_M<\infty$ are threshold parameters. When $s_t=y_{it-j}$ for some variable $i\in\lbrace 1,....,d\rbrace$ and lag $j\in\lbrace 1,....,p\rbrace$, the model is called the self-exciting TVAR model.

The MS-VAR model, introduced by \cite{Krolzig:1997} (see also \citealp{Hamilton:1990}), is quite similar to the GMVAR model of \cite{Kalliovirta+Meitz+Saikkonen:2016} with the exception that the switching-probabilities depend on the preceding regime only. In contrast to STVAR models, regime-switches in the MS-VAR model are discrete and unobserved. Specifically, the MS-VAR model is obtained from Equations~(\ref{eq:stvar1})--(\ref{eq:stvar3}) by replacing the transition weights $\alpha_{m,t}$ with unobservable regime variables $s_{m,t}$, $m=1,...,M$, such that at each $t$, exactly one of them takes the value one and the others take value zero. The regime variable that takes the value one is selected randomly according to the probabilities
\begin{equation}
P(s_{m,t}=1|s_{n,t-1}=1)=p_{mn} \ \text{for all} \ t \ \text{and} \ m,n=1,...,M,
\end{equation}
that satisfy $\sum_{n=1}^Mp_{mn}=1$ for all $m$. In other words, the regime that generates each observation is selected randomly at each $t$ according to the probabilities $p_{mn}$, $m,n=1,...,M$. While the MS-VAR model can flexibly accommodate unobservable discrete regime-switches, unlike our GSTVAR model, it is unable to capture gradual shifts in the dynamics of the data. Moreover, in the same way as in the GMVAR model, it is not possible to recover structural shocks from the fitted model because the regime that generates each observation is not observed. 

Besides nonlinear VAR models, time-varying parameter VAR (TVP-VAR) models (\citealp{Cogley+Sargent:2001}; \citealp{Cogley+Sargent:2005}; \citealp{Primiceri:2005}; \citealp{Koop+Leon-Gonzalez+Strachan:2009}; among others) offer a flexible framework for capturing time-variation in the parameter values that is quite different from our GSTVAR model. The standard deviations of the reduced form innovations are often assumed to evolve as geometric random walks (i.e., they accommodate stochastic volatility), and the rest of the parameter values are typically either constant or evolve as random walks \citep[possibly with mixture innovations, see, e.g.,][]{Koop+Leon-Gonzalez+Strachan:2009}. TVP-VAR models are very capable of capturing gradual as well as abrupt changes in parameter values, but they have the major limitation that the stochastic processes governing the parameters are exogenous to the endogenous variables. Hence, although the parameter estimates can reflect changes in economic conditions, the included variables do not directly influence the evolution of the parameter values, and, thus, potentially important endogenous dynamics of the economy are ignored. In contrast, our GSTVAR model allows the included variables to affect the transition weights at each point of time through the full distribution of the preceding $p$ observations.

\section{A Monte Carlo study}\label{sec:montecarlo}

To assess the properties of the ML estimator, we perform a small scale Monte Carlo study. Because the estimation of our GSTVAR model is computationally demanding, we assume a simple setup 
with two variables ($d=2$), two regimes ($M=2$), and autoregressive order one ($p=1$) in model~(\ref{eq:stvar1})--(\ref{eq:stvar3}). Two different sets of values of the parameter $\theta = (\phi_{1,0},\phi_{2,0},\text{vec}(A_{1,1}),\text{vec}(A_{2,1}),\text{vech}(\Omega_1),\text{vech}(\Omega_2),\alpha_1)$ $(19 \times 1)$ are considered. We refer to the resulting models as as Model~1 and Model~2. In each specification, we use the same values of the covariance matrix and the transition weight parameter, $\text{vech}(\Omega_1)=(0.50, 0.20, 0.30)$, $\text{vech}(\Omega_2)=(0.80, -0.20, 0.50)$, and $\alpha_1 = 0.70$, but the intercepts and AR matrices differ. Hence, we we are able to learn about the importance of the dynamics of the model for the finite sample properties of the estimator. 

The employed values of $\phi_{1,0},\phi_{2,0},\text{vec}(A_{1,1}),\text{vec}(A_{2,1})$ are presented in Table~\ref{tab:monteparams} for Model 1 and Model 2, along with the unconditional means as well as the modulus of the eigenvalues of the companion form AR matrices of the regimes. The latter indicate that Model~1 does not exhibit very high degree of persistence, whereas the AR matrices of Model~2 are somewhat close the boundary of the stationarity region. There is a reasonable amount of variation in the unconditional means across the regimes in both specifications, with less (more) variation for the first (second) component series in Model~1, and vice versa in Model~2. 

\begin{table}[p]
\footnotesize
\renewcommand{\arraystretch}{0.8}
\centering
\begin{tabular}{c c c@{\hspace{3pt}}c c c:c c@{\hspace{3pt}}c c c}
 & \multicolumn{5}{c}{Regime 1} & \multicolumn{5}{c}{Regime 2} \\
 & $\phi_{1,0}$ & \multicolumn{2}{c}{$A_{1,1}$} & $\mu_1$ & $||\boldsymbol{A}_1||$ & $\phi_{2,0}$ &  \multicolumn{2}{c}{$A_{2,1}$} & $\mu_2$ & $||\boldsymbol{A}_2||$ \\ 
\hline\\[-1.5ex]
\multirow{2}{*}{Model 1} & $0.00$ & $0.50$ & $-0.30$           & $-1.43          $ & $0.64$ & $1.50$                          & $-0.10$                   & $-0.20$           & $\phantom{-}0.57$ & $0.37$ \\
                         & $1.00$ & $0.20$ & $\phantom{-}0.70$ & $\phantom{-}2.28$ & $0.64$ & $2.00$ & $\phantom{-}0.30$ & $\phantom{-}0.50$ & $\phantom{-}4.34$ & $0.03$ \\
\hline\\[-1.5ex]
\multirow{2}{*}{Model 2} & $0.00$ & $0.80$ & $-0.55$           & $-2.29$           & $0.97$ & $0.50$                          & $-0.99$                   & $-0.20$           & $-0.19$           & $0.96$\\
                         & $1.00$ & $0.40$ & $\phantom{-}0.90$ & $\phantom{-}0.83$ & $0.97$ & $0.50$ & $\phantom{-}0.30$ & $\phantom{-}0.90$ & $\phantom{-}4.42$ & $0.87$ \\
\end{tabular}
\caption{The parameters values of $\phi_{1,0},\phi_{2,0},A_{1,1}$, and $A_{2,1}$ used for Model~1 and Model~2 in the Monte Carlo study. Also, the corresponding unconditional means of the regimes $\mu_m=(I_d - A_{m,1})^{-1}\phi_{m,0}$, $m=1,2$, and modulus of the eigenvalues of the companion form AR matrices of the regimes $||\boldsymbol{A}_m||$, $m=1,2$.}
\label{tab:monteparams}
\end{table}

\begin{table}[p]
\footnotesize
\renewcommand{\arraystretch}{0.8}
\centering
\begin{tabular}{c c c@{\hspace{3pt}}c c@{\hspace{3pt}}c c@{\hspace{3pt}}c c@{\hspace{3pt}}c c@{\hspace{3pt}}c}

& & \multicolumn{2}{c}{$T=250$} & \multicolumn{2}{c}{$T=500$} & \multicolumn{2}{c}{$T=1000$} & \multicolumn{2}{c}{$T=2000$} & \multicolumn{2}{c}{$T=10000$}\\ 
\hline\\[-1.2ex]
Model 1 & $\phi_{1,0}$ & $\phantom{-}0.04$ & $(0.31)$ & $\phantom{-}0.01$ & $(0.18)$ & $\phantom{-}0.00$ & $(0.12)$ & $\phantom{-}0.00$ & $(0.08)$ & $\phantom{-}0.00$ & $(0.04)$ \\
& & $\phantom{-}0.05$ & $(0.22)$ & $\phantom{-}0.02$ & $(0.15)$ & $\phantom{-}0.01$ & $(0.09)$ & $\phantom{-}0.00$ & $(0.06)$ & $\phantom{-}0.00$ & $(0.02)$ \\
& $\phi_{2,0}$ & $\phantom{-}0.03$ & $(1.23)$ & $\phantom{-}0.07$ & $(0.69)$ & $\phantom{-}0.02$ & $(0.42)$ & $\phantom{-}0.01$ & $(0.30)$ & $\phantom{-}0.00$ & $(0.14)$ \\
& & $\phantom{-}0.12$ & $(0.74)$ & $\phantom{-}0.05$ & $(0.44)$ & $\phantom{-}0.02$ & $(0.30)$ & $\phantom{-}0.01$ & $(0.20)$ & $\phantom{-}0.00$ & $(0.09)$ \\
& $A_{1,1}$ & $-0.02$ & $(0.10)$ & $-0.01$ & $(0.06)$ & $\phantom{-}0.00$ & $(0.04)$ & $\phantom{-}0.00$ & $(0.03)$ & $\phantom{-}0.00$ & $(0.01)$ \\
& & $\phantom{-}0.00$ & $(0.06)$ & $\phantom{-}0.00$ & $(0.04)$ & $\phantom{-}0.00$ & $(0.03)$ & $\phantom{-}0.00$ & $(0.02)$ & $\phantom{-}0.00$ & $(0.01)$ \\
& & $-0.01$ & $(0.09)$ & $-0.01$ & $(0.06)$ & $\phantom{-}0.00$ & $(0.04)$ & $\phantom{-}0.00$ & $(0.03)$ & $\phantom{-}0.00$ & $(0.01)$ \\
& & $-0.02$ & $(0.06)$ & $-0.01$ & $(0.04)$ & $\phantom{-}0.00$ & $(0.03)$ & $\phantom{-}0.00$ & $(0.02)$ & $\phantom{-}0.00$ & $(0.01)$ \\
& $A_{2,1}$ & $-0.02$ & $(0.26)$ & $-0.03$ & $(0.14)$ & $-0.01$ & $(0.08)$ & $\phantom{-}0.00$ & $(0.06)$ & $\phantom{-}0.00$ & $(0.03)$ \\
& & $\phantom{-}0.01$ & $(0.16)$ & $\phantom{-}0.01$ & $(0.09)$ & $\phantom{-}0.00$ & $(0.06)$ & $\phantom{-}0.00$ & $(0.04)$ & $\phantom{-}0.00$ & $(0.02)$ \\
& & $-0.01$ & $(0.25)$ & $-0.01$ & $(0.15)$ & $\phantom{-}0.00$ & $(0.09)$ & $\phantom{-}0.00$ & $(0.06)$ & $\phantom{-}0.00$ & $(0.03)$ \\
& & $-0.04$ & $(0.17)$ & $-0.01$ & $(0.10)$ & $\phantom{-}0.00$ & $(0.06)$ & $\phantom{-}0.00$ & $(0.04)$ & $\phantom{-}0.00$ & $(0.02)$ \\
& $\Omega_{1}$ & $-0.01$ & $(0.07)$ & $\phantom{-}0.00$ & $(0.04)$ & $\phantom{-}0.00$ & $(0.03)$ & $\phantom{-}0.00$ & $(0.02)$ & $\phantom{-}0.00$ & $(0.01)$ \\
& & $-0.01$ & $(0.06)$ & $\phantom{-}0.00$ & $(0.03)$ & $\phantom{-}0.00$ & $(0.02)$ & $\phantom{-}0.00$ & $(0.01)$ & $\phantom{-}0.00$ & $(0.01)$ \\
& & $\phantom{-}0.00$ & $(0.04)$ & $\phantom{-}0.00$ & $(0.03)$ & $\phantom{-}0.00$ & $(0.02)$ & $\phantom{-}0.00$ & $(0.01)$ & $\phantom{-}0.00$ & $(0.01)$ \\
& $\Omega_{2}$ & $-0.05$ & $(0.16)$ & $-0.02$ & $(0.11)$ & $\phantom{-}0.00$ & $(0.07)$ & $\phantom{-}0.00$ & $(0.05)$ & $\phantom{-}0.00$ & $(0.02)$ \\
& & $\phantom{-}0.02$ & $(0.11)$ & $\phantom{-}0.00$ & $(0.07)$ & $\phantom{-}0.00$ & $(0.04)$ & $\phantom{-}0.00$ & $(0.03)$ & $\phantom{-}0.00$ & $(0.01)$ \\
& & $-0.03$ & $(0.10)$ & $-0.01$ & $(0.06)$ & $\phantom{-}0.00$ & $(0.04)$ & $\phantom{-}0.00$ & $(0.03)$ & $\phantom{-}0.00$ & $(0.01)$ \\
& $\alpha_1$ & $\phantom{-}0.01$ & $(0.09)$ & $\phantom{-}0.01$ & $(0.07)$ & $\phantom{-}0.00$ & $(0.05)$ & $\phantom{-}0.00$ & $(0.03)$ & $\phantom{-}0.00$ & $(0.02)$ \\
\hline\\[-1.2ex]
Model 2 & $\phi_{1,0}$ & $-0.08$ & $(0.36)$ & $-0.02$ & $(0.20)$ & $\phantom{-}0.00$ & $(0.08)$ & $\phantom{-}0.00$ & $(0.05)$ & $\phantom{-}0.00$ & $(0.02)$ \\
& & $\phantom{-}0.02$ & $(0.14)$ & $\phantom{-}0.01$ & $(0.09)$ & $\phantom{-}0.00$ & $(0.06)$ & $\phantom{-}0.00$ & $(0.04)$ & $\phantom{-}0.00$ & $(0.02)$ \\
& $\phi_{2,0}$ & $-0.35$ & $(1.10)$ & $-0.18$ & $(0.71)$ & $-0.08$ & $(0.48)$ & $-0.04$ & $(0.32)$ & $-0.02$ & $(0.15)$ \\
& & $\phantom{-}0.16$ & $(0.35)$ & $\phantom{-}0.07$ & $(0.21)$ & $\phantom{-}0.03$ & $(0.13)$ & $\phantom{-}0.01$ & $(0.08)$ & $\phantom{-}0.00$ & $(0.04)$ \\
& $A_{1,1}$ & $-0.11$ & $(0.35)$ & $-0.03$ & $(0.18)$ & $\phantom{-}0.00$ & $(0.08)$ & $\phantom{-}0.00$ & $(0.01)$ & $\phantom{-}0.00$ & $(0.01)$ \\
& & $-0.01$ & $(0.05)$ & $\phantom{-}0.00$ & $(0.03)$ & $\phantom{-}0.00$ & $(0.02)$ & $\phantom{-}0.00$ & $(0.01)$ & $\phantom{-}0.00$ & $(0.01)$ \\
& & $\phantom{-}0.04$ & $(0.19)$ & $\phantom{-}0.01$ & $(0.11)$ & $\phantom{-}0.00$ & $(0.05)$ & $\phantom{-}0.00$ & $(0.03)$ & $\phantom{-}0.00$ & $(0.01)$ \\
& & $-0.02$ & $(0.05)$ & $-0.01$ & $(0.03)$ & $\phantom{-}0.00$ & $(0.02)$ & $\phantom{-}0.00$ & $(0.01)$ & $\phantom{-}0.00$ & $(0.01)$ \\
& $A_{2,1}$ & $\phantom{-}0.19$ & $(0.48)$ & $\phantom{-}0.06$ & $(0.28)$ & $\phantom{-}0.01$ & $(0.11)$ & $\phantom{-}0.00$ & $(0.02)$ & $\phantom{-}0.00$ & $(0.01)$ \\
& & $\phantom{-}0.00$ & $(0.11)$ & $\phantom{-}0.00$ & $(0.06)$ & $\phantom{-}0.00$ & $(0.04)$ & $\phantom{-}0.00$ & $(0.03)$ & $\phantom{-}0.00$ & $(0.01)$ \\
& & $\phantom{-}0.04$ & $(0.26)$ & $\phantom{-}0.02$ & $(0.15)$ & $\phantom{-}0.02$ & $(0.09)$ & $\phantom{-}0.01$ & $(0.06)$ & $\phantom{-}0.00$ & $(0.03)$ \\
& & $-0.03$ & $(0.09)$ & $-0.01$ & $(0.04)$ & $-0.01$ & $(0.03)$ & $\phantom{-}0.00$ & $(0.02)$ & $\phantom{-}0.00$ & $(0.01)$ \\
& $\Omega_{1}$ & $\phantom{-}0.00$ & $(0.10)$ & $\phantom{-}0.00$ & $(0.07)$ & $\phantom{-}0.00$ & $(0.04)$ & $\phantom{-}0.00$ & $(0.03)$ & $\phantom{-}0.00$ & $(0.01)$ \\
& & $-0.03$ & $(0.10)$ & $-0.01$ & $(0.06)$ & $\phantom{-}0.00$ & $(0.03)$ & $\phantom{-}0.00$ & $(0.02)$ & $\phantom{-}0.00$ & $(0.01)$ \\
& & $\phantom{-}0.00$ & $(0.06)$ & $\phantom{-}0.00$ & $(0.04)$ & $\phantom{-}0.00$ & $(0.02)$ & $\phantom{-}0.00$ & $(0.02)$ & $\phantom{-}0.00$ & $(0.01)$ \\
& $\Omega_{2}$ & $-0.05$ & $(0.15)$ & $-0.01$ & $(0.10)$ & $\phantom{-}0.00$ & $(0.07)$ & $\phantom{-}0.00$ & $(0.04)$ & $\phantom{-}0.00$ & $(0.02)$ \\
& & $\phantom{-}0.05$ & $(0.13)$ & $\phantom{-}0.02$ & $(0.08)$ & $\phantom{-}0.00$ & $(0.05)$ & $\phantom{-}0.00$ & $(0.03)$ & $\phantom{-}0.00$ & $(0.01)$ \\
& & $-0.03$ & $(0.10)$ & $-0.01$ & $(0.06)$ & $\phantom{-}0.00$ & $(0.04)$ & $\phantom{-}0.00$ & $(0.03)$ & $\phantom{-}0.00$ & $(0.01)$ \\
& $\alpha_1$ & $-0.03$ & $(0.08)$ & $-0.02$ & $(0.05)$ & $-0.01$ & $(0.03)$ & $\phantom{-}0.00$ & $(0.02)$ & $\phantom{-}0.00$ & $(0.01)$ \\
\end{tabular}
\caption{Results from the Monte Carlo study on the performance of the ML estimator based on $500$ repetitions. The average estimation error is reported for the samples of length $250$, $500$, $1000$, $2000$, and $10000$ in each column first, and next to the biases are the standard deviations of the estimates in parentheses.}
\label{tab:monteresults}
\end{table}

For both model specifications, we generate $500$ samples of length $250,500,1000,2000$, and $10000$. Then, we fit the first-order two-regime GSTVAR model to each of the generated samples based on eight estimation rounds of the two-phase procedure proposed in Section~\ref{sec:estimation}. To estimate the bias of the ML estimator, we calculate the average estimation errors $\hat{\boldsymbol{\theta}} - \boldsymbol{\theta}$ over the $500$ Monte Carlo repetitions, and to estimate the estimation accuracy, we calculate the standard deviations of the estimates.

The results in Table~\ref{tab:monteresults} show that for all the parameters, the bias decreases and estimation accuracy increases substantially as the sample size increases, in line with the consistency of the ML estimator. There is some but mostly small bias when the sample size is small, and particularly the intercept parameters are inaccurately estimated. For most of the parameters, the estimation accuracy seems quite reasonable already in samples of length $500$, however. In the samples of length $2000$ and $10000$, the bias has practically disappears and estimation accuracy is very high for all but some of the intercept parameters, and also the intercept parameters are reasonably well estimated. Finally, there does not appear to be notable differences in estimation accuracy between the two model specifications. 

\section{Details on the empirical application}\label{sec:detailsempapp}
\subsection{Model selection and adequacy of the selected model}
Maximum likelihood (ML) estimation of the models, residual diagnostics, and estimation of the generalized impulse response functions are carried out with the R package sstvars \citep{sstvars} in which the methods introduced in this paper have been implemented. It also contains the dataset studied in the empirical application to facilitate the reproduction of our results. 

\begin{figure}[t]
    \centerline{\includegraphics[width=\textwidth - 2cm]{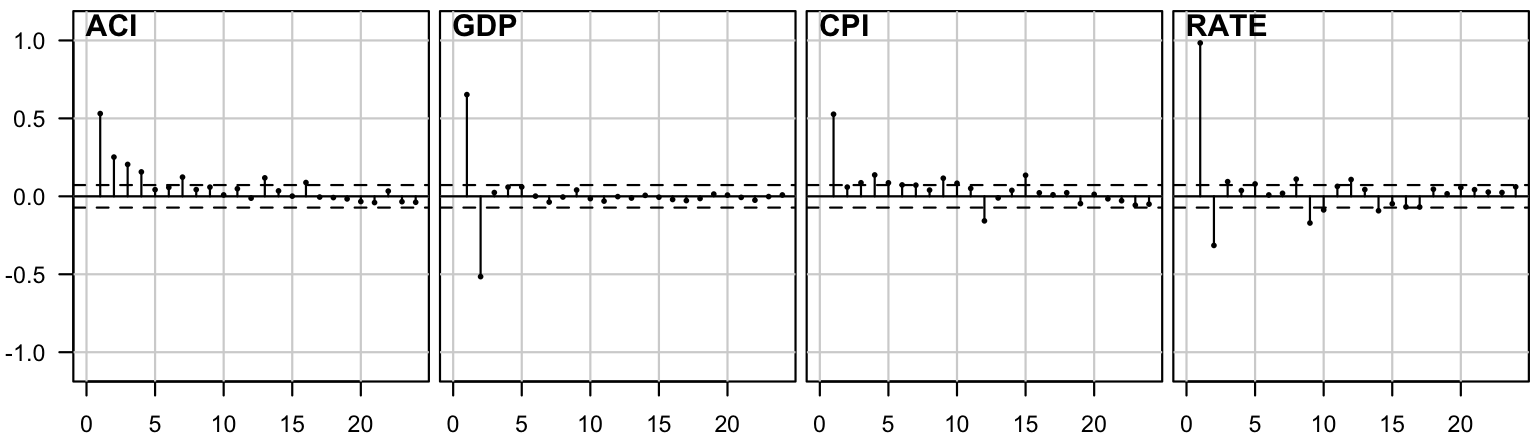}}
    \caption{Partial autocorrelation functions for the first $24$ lags estimated for the monthly U.S. series covering the period from 1961:1 to 2022:3.}
\label{fig:pacf}
\end{figure}

We start model selection with a preliminary examination of the sample partial autocorrelation functions (PACF) of the series, which are presented in Figure~\ref{fig:pacf} for the first $24$ lags. Based on the breaks in the PACFs, an autoregressive order $p=4$ could be reasonable for parsimonious model, and an order smaller than $p=2$ would be clearly insufficient. Therefore, we estimate two-regime GSTVAR models with $p=1,...,5$ and find that the AIC is minimized by the order $p=4$, whereas BIC and HQIC are minimized by the order $p=2$. The values of the information criteria are presented in Table~\ref{tab:ic} (of the paper) together with the maximized log-likelihoods (divided by the number of observations). 

\begin{figure}[t]
    \centerline{\includegraphics[width=\textwidth - 2cm]{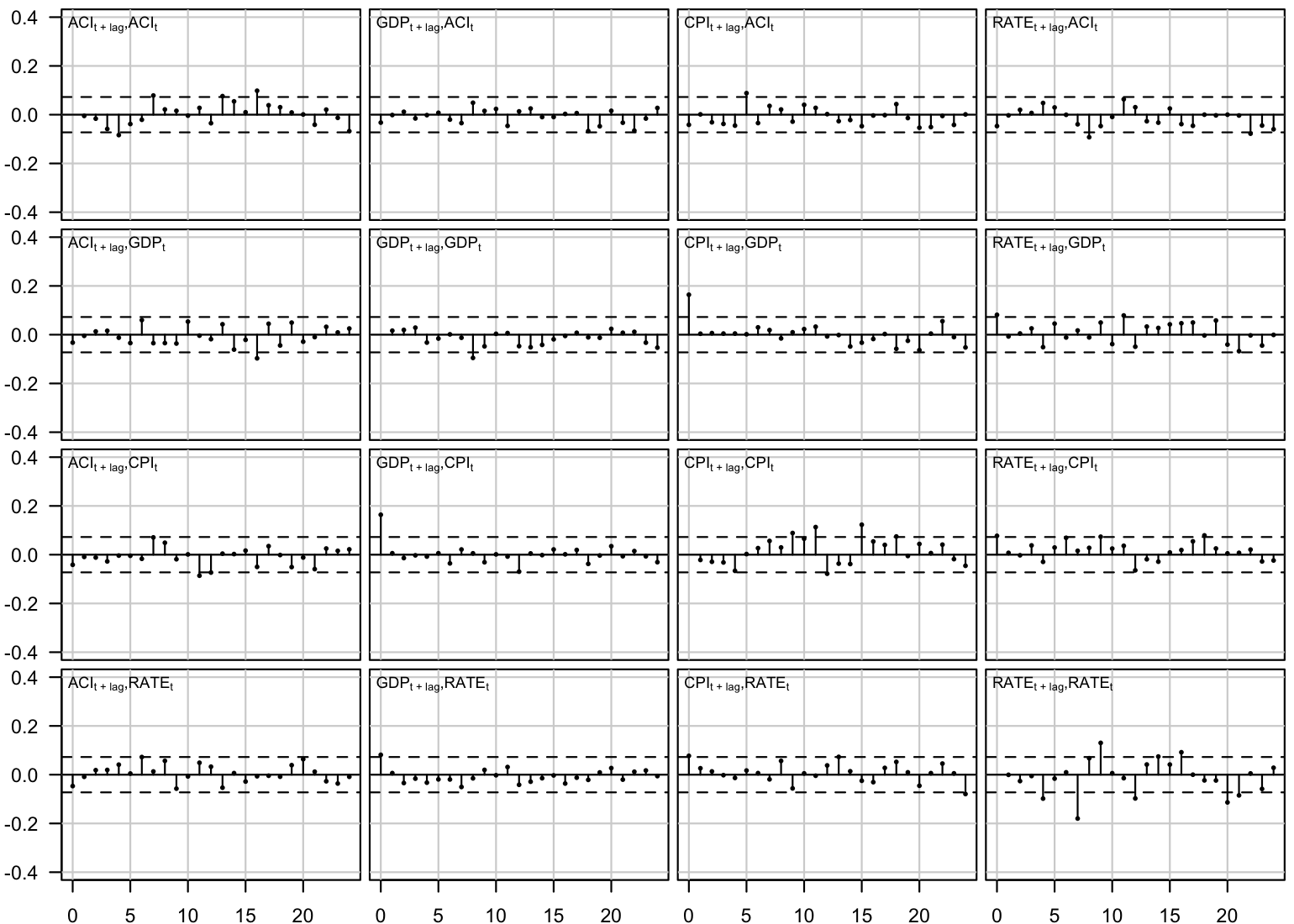}}
    \caption{Auto- and crosscorrelation functions of the residuals of the fitted two-regime GSTVAR $p=4$ model for the lags $0,1,...,24$. The lag zero autocorrelation coefficients are omitted, as they are one by convention. The dashed lines are the $95\%$ bounds $\pm 1.96/\sqrt{T}$ for autocorrelations of IID observations.}
\label{fig:resacf}
\end{figure}

\begin{figure}[t]
    \centerline{\includegraphics[width=\textwidth - 2cm]{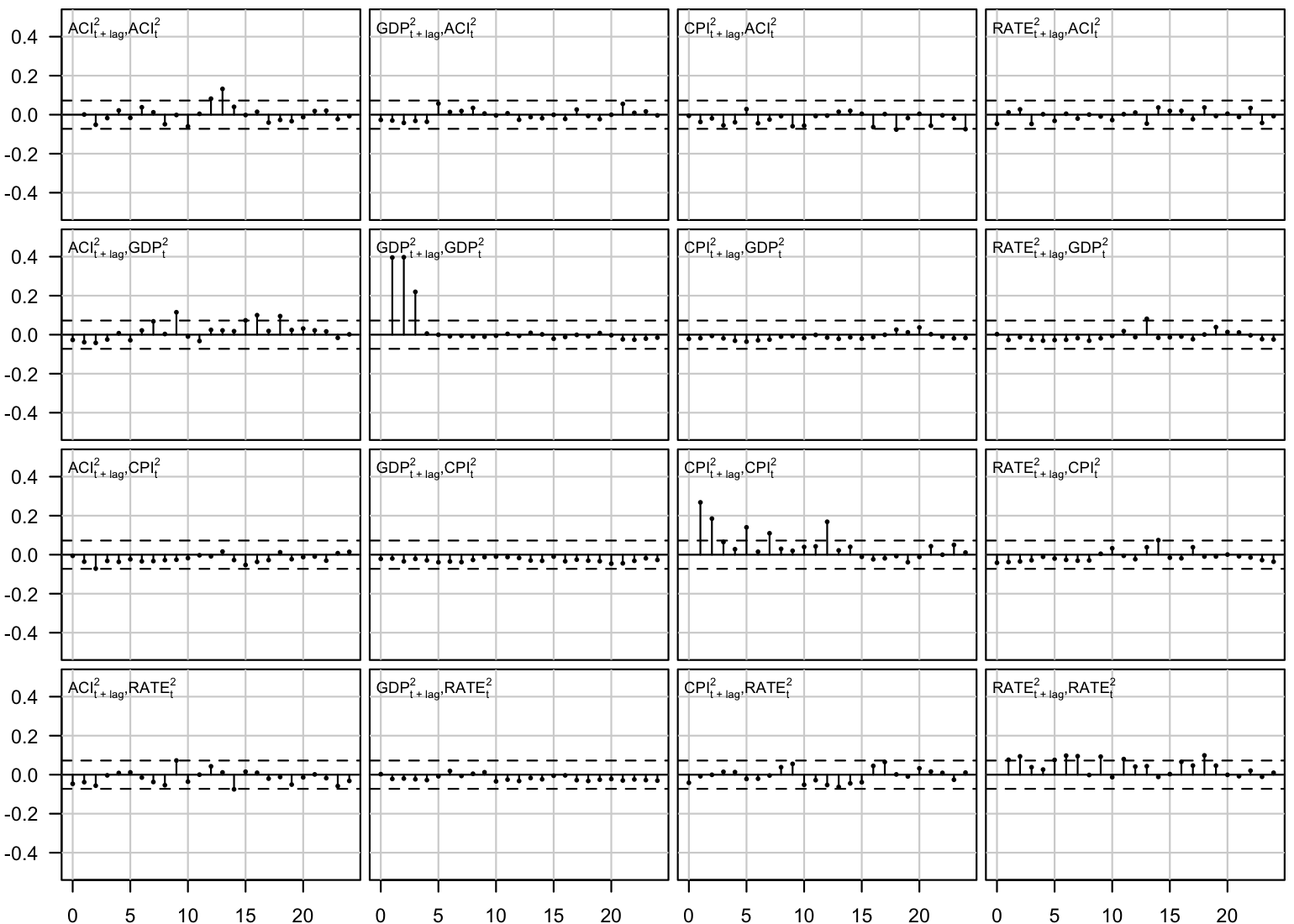}}
    \caption{Auto- and crosscorrelation functions of the squared standardized residuals of the fitted two-regime fourth-order GSTVAR model for the lags $0,1,...,24$. The lag zero autocorrelation coefficients are omitted, as they are one by convention. The dashed lines are the $95\%$ bounds $\pm 1.96/\sqrt{T}$ for autocorrelations of IID observations.}
\label{fig:res2acf}
\end{figure}

\begin{figure}[t]
    \centerline{\includegraphics[width=\textwidth - 2cm]{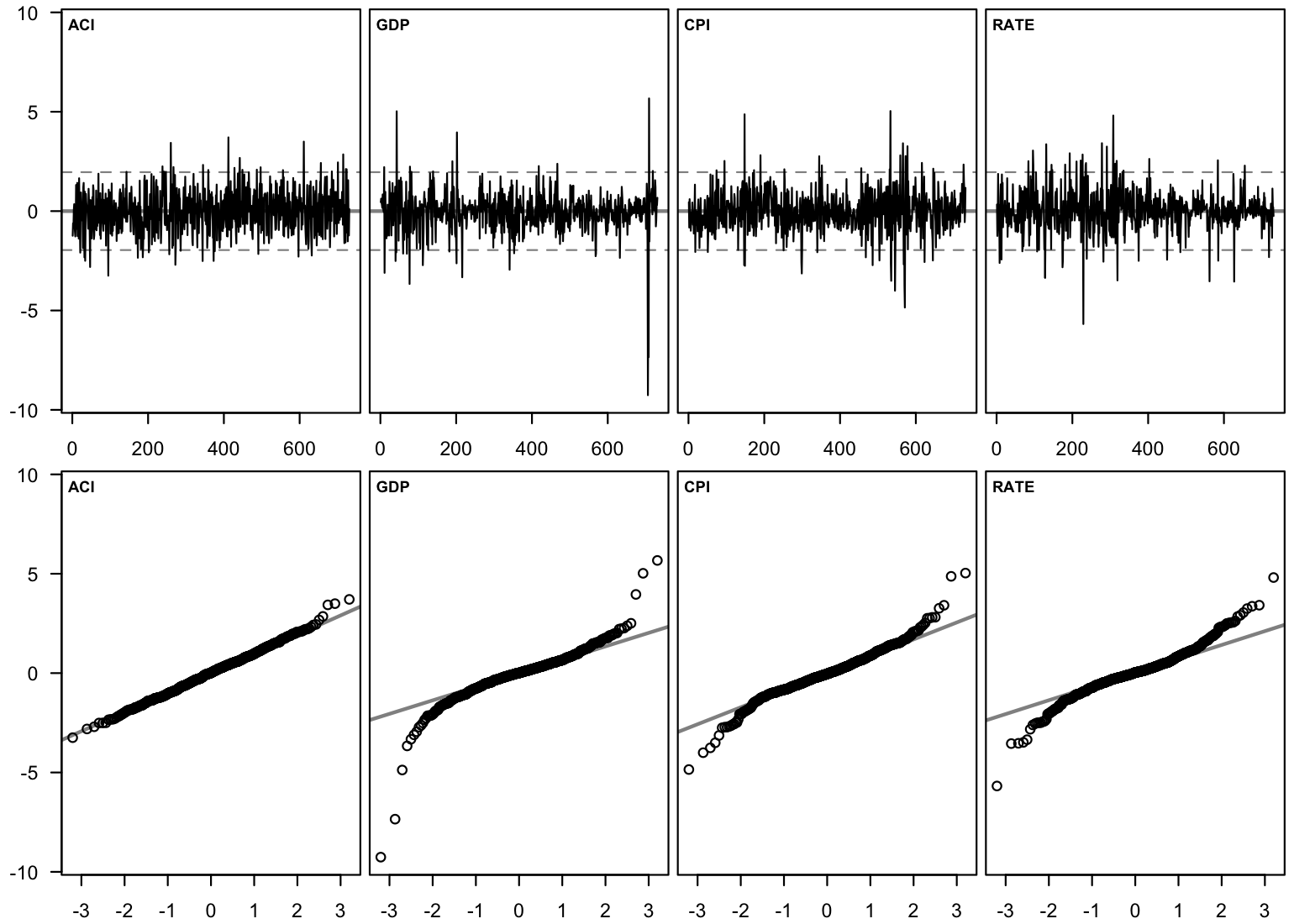}}
    \caption{Standardized residual time series and normal quantile-quantile-plots of the fitted two-regime fourth-order GSTVAR model.}
\label{fig:serqq}
\end{figure}

To study whether our model adequately captures the autocorrelation structure of the data, we depict the sample autocorrelation functions (ACF) and cross-correlation functions (CCF) of the residuals for the first $24$ lags in Figure~\ref{fig:resacf}. As the figure shows, there is not much autocorrelation left in the residuals, but somewhat large correlation coefficient (CC) sticks out at the lag seven in the ACF of the interest rate variable. Moreover, there are somewhat large CCs at the lag zero also in the CCFs between CPI and GDP. Nevertheless, the fitted models seems to capture the autocorrelation structure of the data reasonably well. 

In order to study the model's adequacy to capture the conditional heteroskedasticity of the data, we depict the sample ACFs and CCFs of squared standardized residuals for the first $24$ lags in Figure~\ref{fig:res2acf}. The figure shows that there are several large CCs in the ACFs of the squared standardized residual of GDP and CPI. Otherwise, the CCs are not particularly large. That is, some of the conditional heteroskedasticity of GDP and CPI is not adequately captured, but in our view, the inadequacies are not very severe. 

The series of standardized residuals, presented in the top panels of Figure~\ref{fig:serqq}, also show some remaining heteroskedasticity and several outliers. There is a particularly large negative residual of the GDP growth rate in the beginning of the COVID-19 crisis, when the lockdown of the economy caused a substantial drop in the GDP. However, since the COVID-19 drop was caused by an exceptionally large exogenous shock, a large residual is expected for a correctly specified model. Finally, the normal quantile-quantile-plots, presented in the bottom panel of Figure~\ref{fig:serqq}, show that the marginal distribution of ACI is captured well. The marginal distributions of the residuals of the other variables, however, display excess kurtosis, but they are quite symmetric. Nonetheless, the overall adequacy of the model seems reasonable enough for impulse response analysis.

In our model, the covariance matrix of the errors as well as the autoregressive matrices and the intercept term can vary between the regimes. However, it is possible that the regime shifts are attributed solely to changes in the level or volatility of the series and there is no time-variation in the autoregressive dynamics. Since our GSTVAR model is ergodic stationary, the results of consistency and conventional limiting distribution apply to the ML estimator under conventional high level conditions and standard likelihood based tests can be used to check if the regime shifts are driven only by changes in the volatility regime. The Wald test for the null of constancy of intercepts and AR matrices produces a $p$-value smaller than $10^{-7}$, indicating that the regime-shifts are not solely driven by changes in the volatility regime. The Wald test for the constancy of the AR matrices alone produces a $p$-value smaller than $10^{-5}$. Therefore, we conclude in favor of regime shifts in the level, volatility, as well as the autoregressive dynamics of the series.

\subsection{Robustness checks}\label{sec:robustnesschecks}
\subsubsection{The GSTVAR model fitted to a sub-sample}\label{sec:rob_subsamp}
\begin{figure}
    \centerline{\includegraphics[width=\textwidth - 2cm]{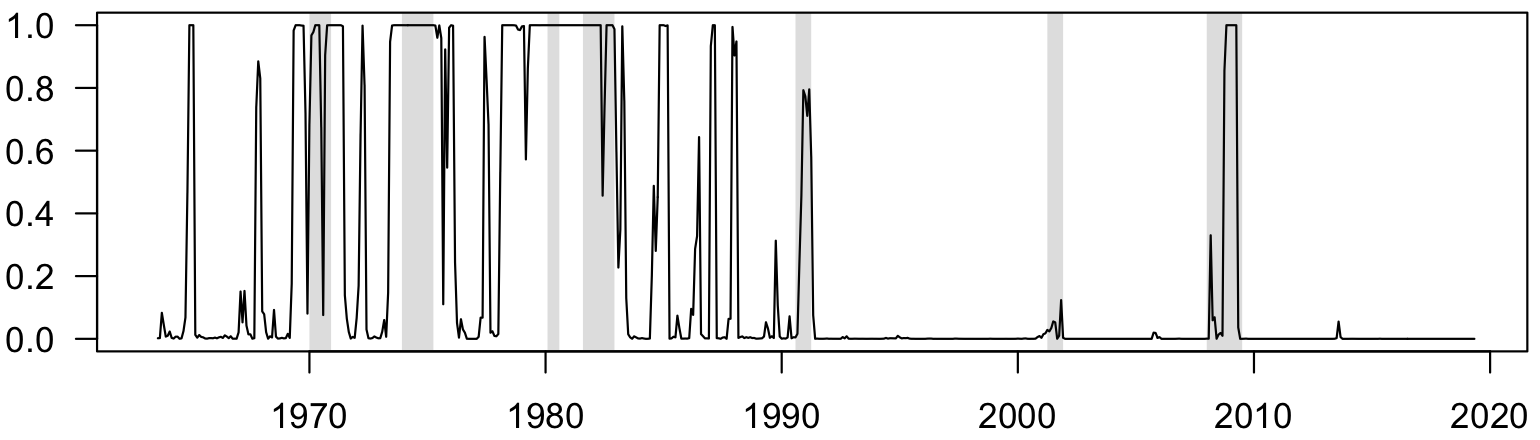}}
    \caption{Estimated transition weights of  Regime 1 for the two-regime fourth-order GSTVAR model fitted to the monthly U.S. data covering the period from 1963:4 to 2019:5. The shaded areas indicate the period of NBER  based U.S. recessions. }
\label{fig:shortdatatransweights}
\end{figure}

As the first robustness check, we consider the subsample period from 1963:4 to 2019:5 studied by \cite{Kim+Matthes+Phan:2022}.\footnote{The ACI and interest rate variable series are similar to theirs, but otherwise our data differ in two main aspects. First, instead of using unemployment rate and year-on-year industrial production index as a measure of real economic activity, we use the monthly GDP growth rate described in Section~\ref{sec:empapp}. Second, instead of using year-on-year CPI inflation rate, we use the monthly CPI inflation growth rate.} We fit two-regime GSTVAR models the subsample and select the autoregressive order $p=4$ based on information criteria and preliminary examinations of partial autocorrelation functions of the series (not shown). The estimated transition weights of our two-regime fourth-order GSTVAR model are presented Figure~\ref{fig:shortdatatransweights} with the shaded areas indicating the periods of NBER based on U.S. recessions. As the transition weights show, the regimes are quite similar to the benchmark specification: Regime~1 mainly dominates in the volatile period of 1960s, 1970s, and 1980s, but obtains obtains large weights also in the 1990s recession and the Financial crisis. Conversely, Regime~2 dominates in the less volatile periods. 

\begin{figure}[!t]
    \centerline{\includegraphics[width=\textwidth - 2cm]{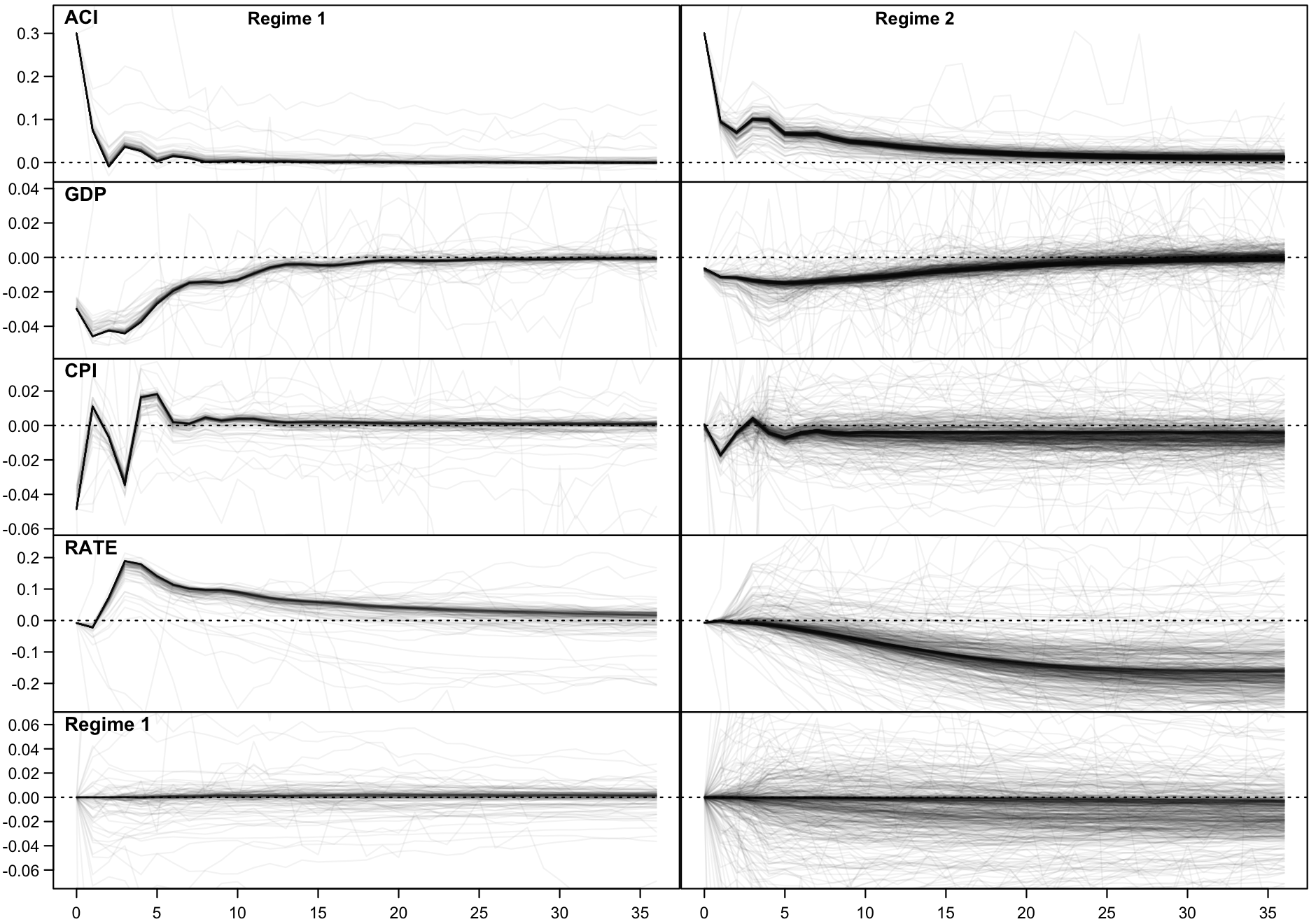}}
    \caption{Generalized impulse response functions to the identified ACI shock $h=0,1,....,36$ months ahead, based on the data from from 1963:4 to 2019:5. From the top to bottom, the responses of ACI, GDP growth rate, CPI growth rate, the interest rate variable, and transition weights of Regime~1 are depicted in the rows. The left and right columns show the responses conditional on Regime~1 and Regime~2 dominating when the shock arrives, respectively. All GIRFs have been scaled so that instantaneous effect of the ACI index is $0.3$. The results are based on $2500$ Monte Carlo repetitions for each history (see Appendix~\ref{sec:montecarlo_girf} for details).}
\label{fig:shortdatagirfplot}
\end{figure}

The GIRFs estimated for the model are presented in Figure~\ref{fig:shortdatagirfplot}, which is similar to Figure~\ref{fig:girfplot} of the paper. The GIRFs are also mostly similar to our main specification utilizing the full sample period from 1961:1 to 2022:3. In both of the regimes, a positive ACI shock decreases GDP and consumer prices, but the effects are stronger in Regime 1. A notable difference to our main specification is that a positive ACI shock increases the interest rate in Regime~1, but decreases it in Regime~2.

\subsubsection{The GMVAR model and a linear SVAR model}\label{sec:rob_gmvar}
\begin{figure}
    \centerline{\includegraphics[width=\textwidth - 2cm]{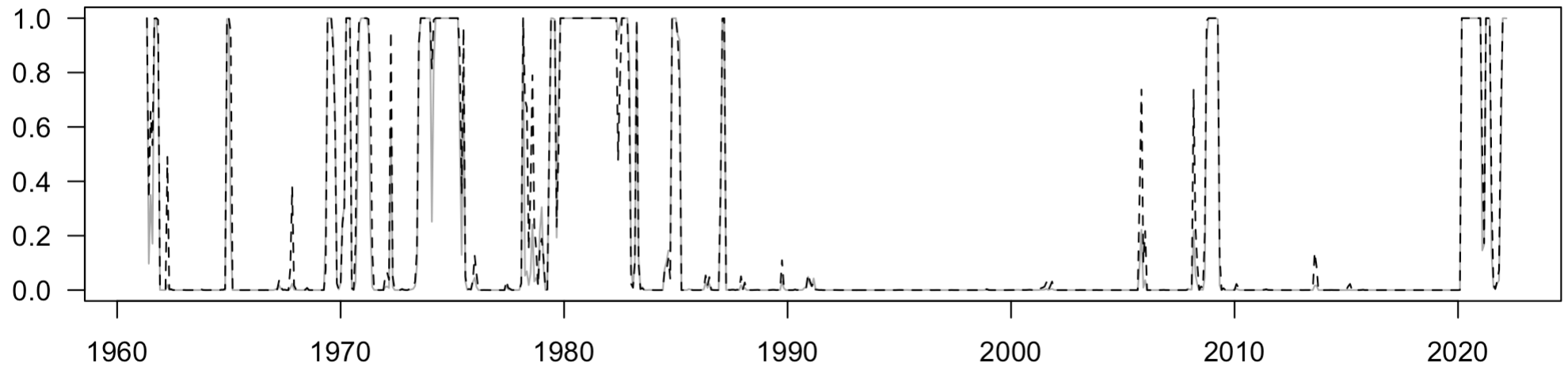}}
    \caption{Estimated mixing weights of Regime 1 of the two-regime fourth-order GMVAR model fitted to the full sample period (black dashed line) together with the estimated transition weights of Regime 1 of our GSTVAR model (grey solid line). }
\label{fig:rob_weightplot}
\end{figure}

For comparison, we also fit the two-regime fourth-order GMVAR model \citep{Kalliovirta+Meitz+Saikkonen:2016} as well as the recursively identified linear Gaussian SVAR model (with autoregressive order $p=9$ based on the AIC) to the full sample period. The estimated mixing weight of the GMVAR model, presented in Figure~\ref{fig:rob_weightplot} (black dashed line), is very similar to the estimated transition weights of our GSTVAR model. Hence, the regimes of the GMVAR model and our GSTVAR model have similar interpretations. 

\begin{figure}
    \centerline{\includegraphics[width=\textwidth - 2cm]{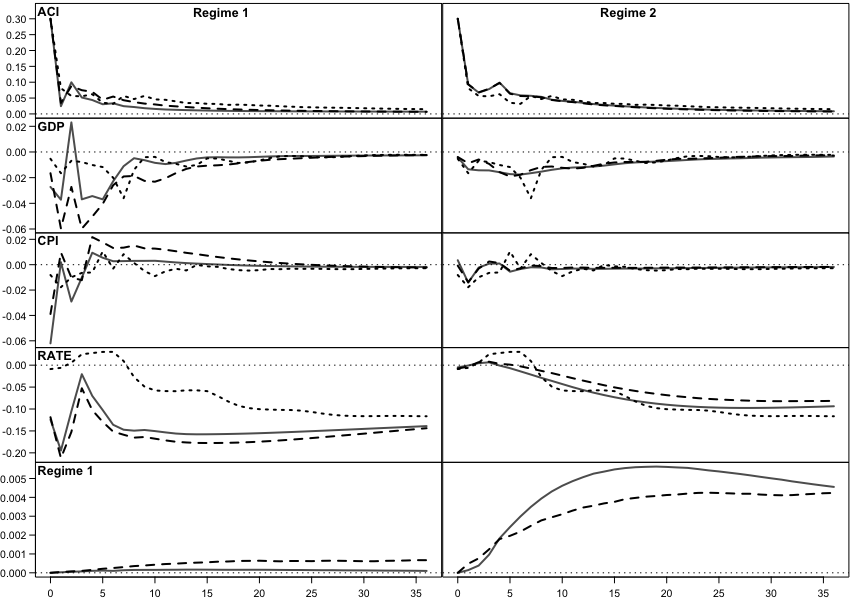}}
    \caption{The (generalized) impulse response functions to the identified ACI shock $h=0,1,....,36$ months ahead in the GSTVAR model (grey solid line), the GMVAR model (black dashed line), and the linear Gaussian SVAR model (black dotted line). From top to bottom, the responses of ACI, GDP growth rate, CPI growth rate, the interest rate variable, and transition weights of Regime 1 are depicted in the rows. The left and right columns show the responses to a positive one-standard-error shock with histories generated from the stationary distribution of Regime~1 and Regime~2, respectively. All impulse responses have been scaled so that the instantaneous change in the ACI index is $0.3$. The GIRFs are based on $2500$ draws of initial values (for details of the algorithm, see Appendix~\ref{sec:montecarlo_girf}). For the linear SVAR, the conventional IRFs are depicted in both columns.}
\label{fig:rob_girfplot}
\end{figure}

Figure~\ref{fig:rob_girfplot} presents the GIRFs to a positive one-standard-error ACI shock in the GMVAR model (black dashed line) with the histories generated from the stationary distribution of each regime. Only the average GIRFs over the histories are depicted for clarity. For easy comparison, also the average GIRFs produced in a similar manner for our GSTVAR model are depicted (grey solid line). The GIRFs based on the GMVAR model are quite similar to those of our GSTVAR model in the sense that in both regimes, GDP and the interest rate decrease and there is a short-term decrease in inflation rate. Like in our GSTVAR model, the responses of GDP, inflation rate, and interest rate are clearly stronger in the more turbulent Regime~1, and the responses of the mixing weights are very weak in both regimes. 

To study possible asymmetries in the effects of the ACI shock with respect to its sign and size, we calculate the GIRFs also for a negative one-standard-error ACI shock as well as for positive and negative two-standard-error ACI shocks. However, marked asymmetries in neither the GMVAR nor the GSTVAR model can be found. Therefore, to save space, the GIRFs for negative or two-standard-error ACI shock are not presented.

Figure~\ref{fig:rob_girfplot} also presents the impulse response functions (IRF) to the ACI shock in the linear SVAR model (black dotted line, depicted in both of the columns of Figure~\ref{fig:rob_girfplot}). Like in the GSTVAR model, the IRFs show a decrease in GDP, CPI and interest rate in response to a positive ACI shock, but the linear model is, of course, incapable of capturing the differences across the regimes. 

\end{appendices}

\end{document}